\documentclass[lettersize,journal]{IEEEtran}
\usepackage{amsmath,amsfonts}
\usepackage{algorithmic}
\usepackage{algorithm}
\usepackage{array}
\usepackage{textcomp}
\usepackage{stfloats}
\usepackage{url}
\usepackage{verbatim}
\usepackage{graphicx}
\usepackage{cite}
\usepackage{subfloat}
\usepackage{booktabs}
\usepackage{cleveref}
\usepackage{utfsym}
\usepackage{lineno}
\begin{document}
\title{DRL-Based Robust Multi-Timescale Anti-Jamming Approaches under State Uncertainty}

\author{Haoqin Zhao, Zan Li, \IEEEmembership{Fellow}, \IEEEmembership{IEEE}, Jiangbo Si, \IEEEmembership{Senior Member}, \IEEEmembership{IEEE}, Rui Huang, Hang Hu,
	
Tony Q.S. Quek, \IEEEmembership{Fellow}, \IEEEmembership{IEEE}, and Naofal Al-Dhahir, \IEEEmembership{Fellow}, \IEEEmembership{IEEE}
\thanks{This work was supported in part by the National Natural Science Foundation of China under Grant 62425103; in part by Shaanxi Province Natural Science Basic Research Program under Grant 2024JC-YBMS-514.}
\thanks{Haoqin Zhao, Zan Li, Jiangbo Si, Rui Huang and Hang Hu are with the State Key Laboratory of Integrated Services Networks, School of Telecommunication Engineering, Xidian University, Xi'an 710071, China (e-mail: hqzhao@stu.xidian.edu.cn; zanli@xidian.edu.cn; jbsi@xidian.edu.cn; ruihuang@stu.xidian.edu.cn; xd\_huhang@126.com).

Tony Q.S. Quek is with the Singapore University of Technology and Design, Singapore 487372 (e-mail: tonyquek@sutd.edu.sg)

Naofal Al-Dhahir is with the Department of Electrical and Computer Engineering. The University of Texas at Dallas, Richardson, TX 75080 USA (e-mail: aldhahir@utdallas.edu).
}}

\markboth{IEEE TRANSACTIONS ON COGNITIVE COMMUNICATIONS AND NETWORKING,~Vol.~X, No.~X, XX~2025}%
{Zhao \MakeLowercase{\textit{et al.}}: DRL-Based Robust Multi-Timescale Anti-Jamming Approaches under State Uncertainty}

\IEEEpubid{0000--0000/00\$00.00~\copyright~2025 IEEE}

\maketitle

\begin{abstract}
Owing to the openness of wireless channels, wireless communication systems are highly susceptible to malicious jamming. Most existing anti-jamming methods rely on the assumption of accurate sensing and optimize parameters on a single timescale. However, such methods overlook two practical issues: mismatched execution latencies across heterogeneous actions and measurement errors caused by sensor imperfections. Especially for deep reinforcement learning (DRL)-based methods, the inherent sensitivity of neural networks implies that even minor perturbations in the input can mislead the agent into choosing suboptimal actions, with potentially severe consequences. To ensure reliable wireless transmission, we establish a multi-timescale decision model that incorporates state uncertainty. Subsequently, we propose two robust schemes that sustain performance under bounded sensing errors. First, a Projected Gradient Descent–assisted Double Deep Q-Network (PGD-DDQN) algorithm is designed, which derives worst-case perturbations under a norm-bounded error model and applies PGD during training for robust optimization. Second, a Nonlinear Q-Compression DDQN (NQC-DDQN) algorithm introduces a nonlinear compression mechanism that adaptively contracts Q-value ranges to eliminate action aliasing. Simulation results indicate that, compared with the perfect-sensing baseline, the proposed algorithms show only minor degradation in anti-jamming performance while maintaining robustness under various perturbations, thereby validating their practicality in imperfect sensing conditions.
\end{abstract}


\section{Introduction}
\IEEEPARstart{N}{owadays}, as one of the most crucial communication technologies, wireless communication has achieved seamless connectivity from terrestrial to space, having permeated every aspect of modern life \cite{ref1, ref2, ref3_1, ref3_2}. However, the openness of wireless channels makes signals susceptible to malicious jamming, where adversaries disrupt legitimate transmissions by emitting high-power signals on the same frequency band \cite{ref3, ref4}. In both civilian and military applications, the anti-jamming capability of wireless systems directly determines their operational reliability in complex electromagnetic environments. Although frequency-hopping and direct-sequence spread spectrum techniques have achieved widespread application \cite{ref5, ref6, ref7}, the continuous escalation of jamming techniques coupled with rapid advancements in signal processing and artificial intelligence, ensuring highly reliable transmission through intelligent anti-jamming schemes remains a critical issue for the future development of wireless communication technologies.

\subsection{Prior Works}
DRL has proven effective for anti-jamming communications, enabling autonomous decision-making in unknown and dynamic environments \cite{ref8, ref9, ref10, ref11, ref12}. Current works mainly employ single- or multi-dimensional schemes, achieving notable performance gains in mitigating jamming \cite{ref13, ref14, ref15, ref16, ref17, ref18}.

In single-domain anti-jamming schemes, research efforts primarily focus on frequency-domain avoidance and power-domain adaptive adjustment. To address conventional sweep jamming, the study in \cite{ref13} proposes a wideband anti-jamming hopping communication deep Q-network (WAH-DQN) algorithm, which can significantly enhance anti-jamming performance. Furthermore, targeting more advanced active jamming, an enhanced multi-action deep recurrent Q-network algorithm is developed \cite{ref14}. This approach introduces the frequency set to generate frequency hopping sequences under given states, while producing non-sequential hopping sequences to counter tracking jamming. Additionally, transmission power reduction could be implemented to evade detection by active jammers \cite{ref15}. When multi-agent coexist, mutual jamming can be minimized through optimized power allocation \cite{ref16}. 
\IEEEpubidadjcol

When single-domain anti-jamming performance fails to meet requirements, multi-domain actions can be leveraged to achieve performance gains. A common technical approach involves constructing a dual action space comprising frequency and power parameters \cite{ref17}. This framework employs a dual-network architecture for parallel decision-making and incorporates an action feedback mechanism, which enhances system throughput while accounting for energy consumption and frequency switching overhead. The Broad Q-network (BQN) merges a specialized flattened neural network, broad learning system, with the DRL framework, and achieves faster acquisition of anti-jamming strategies in frequency, power, and rate domains. This architecture achieves faster data transmission rates in unknown dynamic environments \cite{ref18}. 

As jamming technologies advance, interference becomes more dynamic. However, the aforementioned methods allocate resources over a single timescale and overlook the mismatch in decision latencies among heterogeneous actions. For instance, frequency switching incurs additional overheads such as queuing delays and transceiver synchronization \cite{ref20}, whereas power and modulation adjustments can be performed more rapidly with lower cost. This latency mismatch may reduce the adaptability of the policy to rapidly varying jamming.

More critically, these methods assume perfect sensing, while practical devices suffer bounded measurement errors due to hardware imperfections \cite{ref20_1}, thereby introducing uncertainty into the state. Most existing studies convexify the original non-convex problem and solve it using convex optimization tools, with a primary focus on addressing imperfect channel state information (CSI) \cite{ref20_1_1,ref20_1_2, ref20_1_3}. They typically formulate the optimization problem as a max–min framework to maximize system performance under the worst-case scenario. Semidefinite relaxation (SDR) is employed to relax the rank-one constraints, while the S-procedure is adopted to approximate the semi-infinite inequality constraints. The resulting problem can then be iteratively solved using successive convex approximation (SCA) to enhance the worst-case performance.

However, robust optimization for DRL has been rarely investigated. Since DRL employs deep neural networks (DNNs) as function approximators for policy learning, the intrinsic vulnerability of neural architectures means that well-trained DNNs can suffer significant performance degradation under minor input perturbations \cite{ref21, ref22, ref23, ref24, ref25}. 

\subsection{Motivation and Contributions}
Focusing on the real problems existing in practical applications: (1) the lag in decision-making efficacy under rapidly varying jamming, and (2) the performance degradation of trained networks caused by sensing errors.
We incorporate state uncertainty into the multi-timescale model and propose two robust anti-jamming schemes. These schemes rectify the neural network during the training phase to sustain anti-jamming performance by tolerating state uncertainty. The main contributions of this work are summarized as follows

\begin{itemize}
	\item Different from existing literature, we investigate a more practical scenario where the electromagnetic environment sensing results are inaccurate, and consider the decision-making latency induced by the rapid jamming. To resolve this, we propose a multi-timescale model under bounded sensing errors, which optimizes anti-jamming actions across distinct timescales. Consequently, the anti-jamming problem is formulated as an uncertain multi-timescale Markov decision process (UM-MDP).
	
	\item To mitigate unstable neural network outputs and performance degradation induced by sensing errors, we propose a PGD-DDQN algorithm. Specifically, the existence of worst-case perturbations is derived, and these conditions are simulated using the PGD method. Then the optimal actions derived from true states are utilized as supervised labels and integrated as regularization terms into the loss function. This approach enhances the lower bound performance in uncertain states.
	
	\item To guarantee the invariance of the output policy under perturbed states, we propose a NQC-DDQN algorithm. Leveraging the property of neural networks where bounded inputs result in bounded outputs, this method constructs a nonlinear compression mechanism. Through adaptively correcting the Q-value intervals of different actions, it eliminates the Q-value aliasing between the optimal action and other actions, thereby establishing immunity to sensing errors.
	
	\item Simulation results first confirm the necessity of multi-timescale decision-making: the proposed scheme achieves a 93.12\% throughput gain compared to single-timescale method. Furthermore, under state uncertainty, the PGD-DDQN algorithm exhibits relatively low volatility, while the NQC-DDQN algorithm maintains complete stability within the predefined tolerance bound. Overall, the proposed algorithms achieve an effective balance between anti-jamming performance and robustness.
	
\end{itemize}

\subsection{Organization}
The rest of this paper is structured as follows. Section \uppercase\expandafter{\romannumeral2} presents the system model and optimization problem formulation. Section \uppercase\expandafter{\romannumeral3} proposes the PGD-DDQN algorithm, and Section \uppercase\expandafter{\romannumeral4} elaborates on the NQC-DDQN algorithm in detail. Section \uppercase\expandafter{\romannumeral5} analyzes the simulation results. Section \uppercase\expandafter{\romannumeral6} summarizes this paper.

\section{System Model}
\subsection{Communication scenario}
\begin{figure}[htbp]
	\centering
	\includegraphics[width=3.2in]{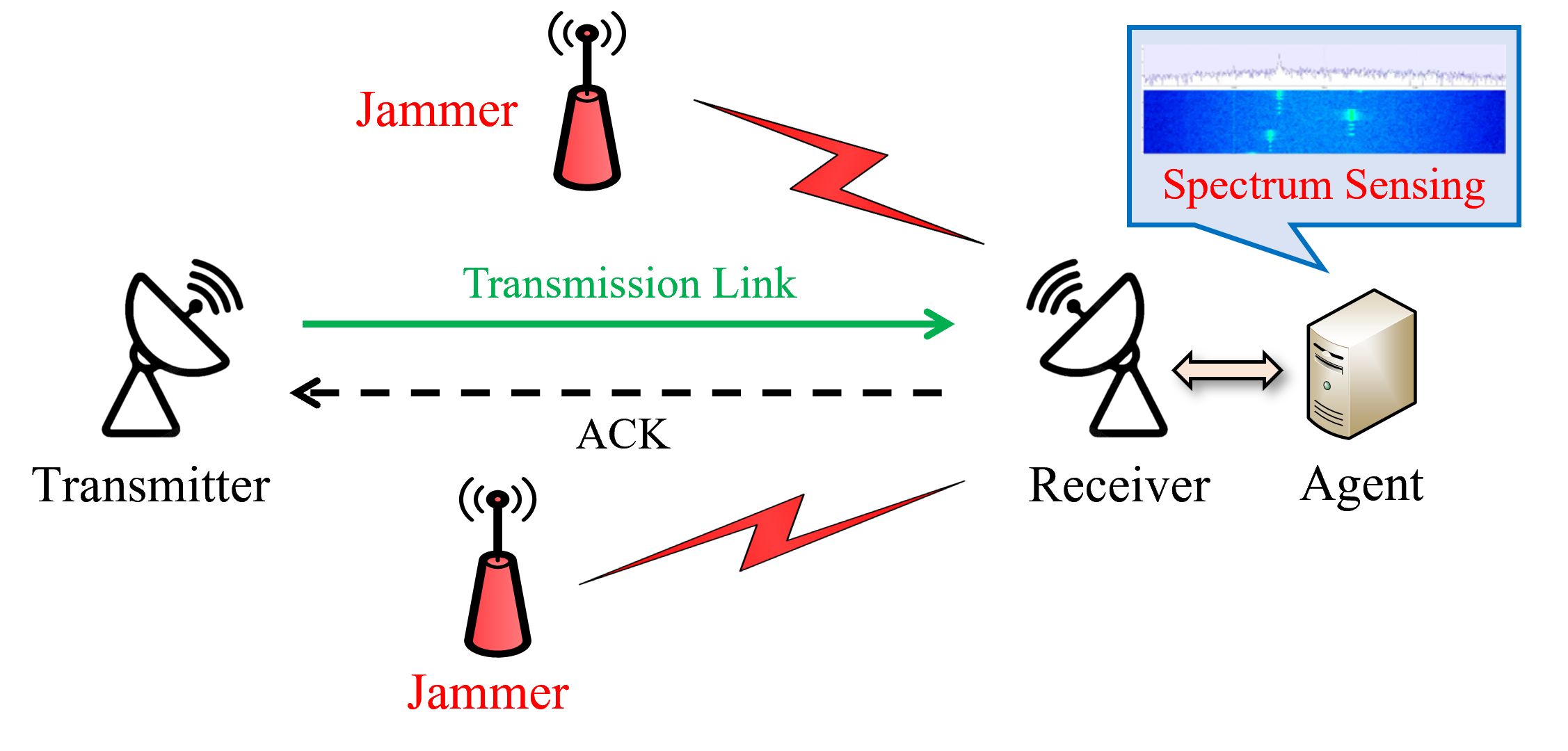}
	\caption{Communication System Model.}
	\label{fig_model}
\end{figure}

We consider a link-level anti-jamming communication scenario, as shown in Fig. \ref{fig_model}, consisting of a transceiver pair and $I$ unknown jammers. These jammers disrupt the receiver by emitting targeted interference signals but, constrained by limited power, cannot perform full-band jamming \cite{ref26}. To counter this, an agent at the receiver performs real-time wideband spectrum sensing and generates anti-jamming schemes in electromagnetic environments, which are then delivered to the transmitter via a low-rate control link for execution.

To enhance anti-jamming performance, we optimize multi-domain parameters including frequency, power, and modulation. The spectrum is divided into $n$ orthogonal subchannels $F=\{ {f_1},{f_2}, \ldots ,{f_n}\}$, each with bandwidth $B$. The available power levels and modulation schemes are defined as $P = \{ {p_1},{p_2}, \ldots ,{p_u}\}$ and $V = \{ {v_1},{v_2}, \ldots ,{v_z}\}$, respectively. Note that even for a single transceiver, the transmit power is dynamically adjusted rather than fixed at its maximum level, in order to avoid detection by reactive cognitive jammers.
\begin{figure}[htbp]
	\centering
	\vspace{-0.1cm}  
	\includegraphics[width=2.8in]{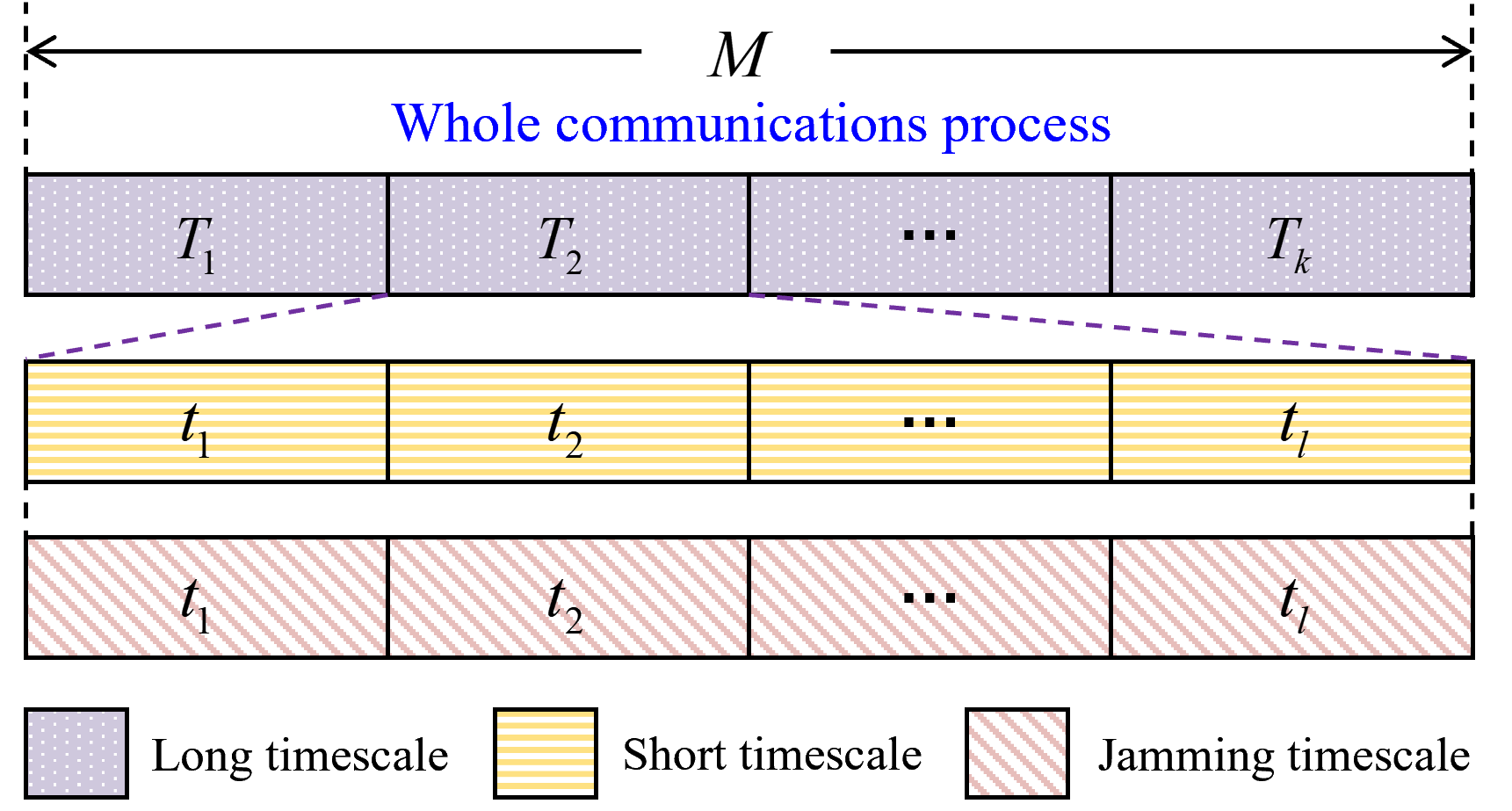}
	\caption{Definition of multi-timescale communication structure.}
	\label{fig_timeslot}
\end{figure}

Considering the different response times of the three variables, a multi-timescale communication structure is designed, as shown in Fig. \ref{fig_timeslot}. The total duration $M$ is divided into $k$ long-timescale $T$ ($M = k \cdot T$), where the frequency scheme remains fixed. Each $T$ is further divided into $l$ short-timescale $t$ for finer control. Moreover, given the shorter latency of jammer's decision cycles, the jamming timeslots are aligned with short-timescales for ease of discussion. Accordingly, frequency is updated in $T$, while power and modulation are optimized within $t$. Assuming a block fading model and the channel gain remains constant in $t$ \cite{ref27}, denoted as
\begin{equation}
	\label{gain}
	g = {(\frac{d}{{{d_0}}})^{ - \tau }} \cdot {\left| h \right|^2},
\end{equation}
where ${d_0}$ is the reference distance, $d$ denotes the distance from the transmitter or jammer to the receiver, $\tau$ is the path-loss exponent, and $h$ indicates the channel coefficients under small-scale Rayleigh fading, and $h \sim \mathcal{CN}(0,1)$.

\subsection{Problem Formulation}
The system’s anti-jamming performance is evaluated by the sum throughput over $M$ \cite{ref28}. At $t_l$, when the transmitter operates on channel $f_n$ with power $p_u$, the received power is ${p_r} = {p_u} g_{tr,{t_l}}^{{f_n}}$, where $g_{tr,{t_l}}^{{f_n}}$ is the transmitter–receiver channel gain. The jamming power is ${\hat p_j} = {p_{j,tr}} g_{j,{t_l}}^{{f_j}} \delta({f_j}={f_n})$, where $p_{j,tr}$ is the jammer’s emission power, $g_{j,{t_l}}^{{f_j}}$ is the jamming channel gain, and $\delta(\cdot)=1$ when $f_j=f_n$, otherwise $\delta(\cdot)=0$. Furthermore, we consider a more realistic setting in which the agent’s sensing results are influenced by measurement errors, resulting in uncertain state observations. The transmit power of $I$ jammers is modeled as
\begin{equation}
	\label{jam_power}
	{p_{j,i}} = {\hat p_{j,i}} + \Delta {p_{j,i}},\forall i \in I,
\end{equation}
where $\Delta {p_{j,i}}$ is the uncertain error of $i$-th jammer. Since the error of the actual detection device is usually bounded, the detection error model can be formulated as
\begin{equation}
	\label{bound_model}
	{\left\| {\Delta {p_{j,i}}} \right\|_2} \le {\varepsilon},\forall i \in I,
\end{equation}
where ${\varepsilon}$ is the radius of the uncertainty region known by the agent. Thus, for the receiver, the signal to jamming plus noise ratio (SJNR) in $t$ is expressed as
\begin{equation}
	\label{SINR}
	{\beta _t} = \frac{{{p_r}}}{{\sum\nolimits_{i = 1}^I {{p_{j,i}}}  + {\sigma ^2} }},
\end{equation}
where ${{\sigma ^2}}$ denotes the noise power at the receiver. 

Under uncertain jamming power, the objective is to maximize throughput through multi-timescale optimization of frequency, power, and modulation while ensuring robustness. The optimization problem is formulated as

\begin{subequations}\label{objective}
	\setlength{\abovedisplayskip}{2pt}
	\begin{align}
		\mathop {Max}\limits_{{f_n},{p_u},{v_z}} \quad &  \sum\limits_{T = 1}^k {\sum\limits_{t = 1}^l {\mu  \cdot \psi ({v_z},{\beta _t})} }  \cdot B \cdot {\log _2}(1 + {\beta _t})  \\  
		\text{s.t.} \quad & \eqref{bound_model},  \\  
		& \mu  \in \{ 0,1\},  \\  
		& \psi ({v_z},{\beta _t}) \in [0,1], \\  
		& f_n \in \{ {f_1},{f_2}, \ldots ,{f_n}\},  \\  
		& p_u \in \{ {p_1},{p_2}, \ldots ,{p_u}\},  \\
		& v_z \in \{ {v_1},{v_2}, \ldots ,{v_z}\},  	   
	\end{align}
\end{subequations}
where the constraint (5b) bounds each jammer’s power perturbation within an $\varepsilon$-radius uncertainty region. Constraint (5c) denotes the successful reception indicator, which equals 1 only when the throughput exceeds the required threshold ${\mu_{\mathrm{th}}}$. Constraint (5d) characterizes the impact of modulation on throughput, where $\psi(v_z,\beta_t)$ serves as a scaling factor whose computation is detailed in Appendix A. (5e), (5f), (5g) denote the set of feasible actions, which are implemented by the agent.

\subsection{Uncertain State Multi-timescale Markov Decision Process}
\begin{figure*}[htbp]
	\setlength{\abovecaptionskip}{0.1cm}   
	\centering
	\includegraphics[width=6.46in]{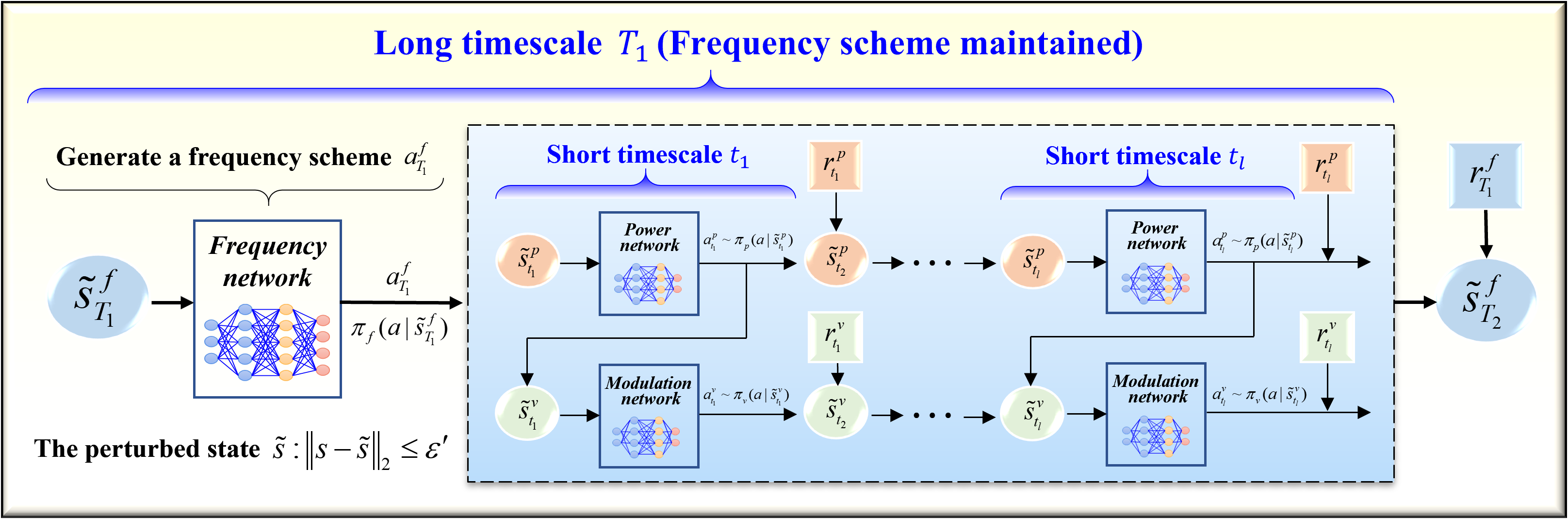}
	\caption{The proposed uncertain state multi-timescale Markov decision process (UM-MDP), which illustrates in detail the decision process within $T_1$.}
	\label{UM_MDP}
\end{figure*}

The anti-jamming communication has been rigorously formulated as a Markov decision process \cite{ref29, ref30}, denoted by a tuple $(S,A,P,R,\gamma )$. Given the bounded error in sensing device, we introduce $\tilde S \in {B_{\varepsilon'} }(S)$ to indicate the perturbed state, where ${B_{\varepsilon'}}(S)$ denotes a ${\ell _2}$-norm ball centered at $S$ with radius $\varepsilon' $. This leads to an augmented tuple $(S, \tilde S, A,P,R,\gamma )$, where $S$ is the state space, which is the underlying state in the environment. $A$ stands for the action space, $P:S \times A \to \Delta S$ defines the state transition probability, note that the $\tilde S$ is simply a perturbation of the agent's observations, the next state still transits from the real state $S$. The reward function is $R$ and discount factor $\gamma  \in [0,1)$. In a deterministic state $s$, actions are taken by the agent following policy $\pi (a|s)$. However, when the agent observes a perturbed state $\tilde s$, the action from the same policy $\pi (a|\tilde s)$ may be sub-optimal, and result in lower reward. Furthermore, in (5a), the optimization variables are optimized at different timescales and coupled with each other, for this reason we propose the UM-MDP, as shown in Fig. \ref{UM_MDP}. A phase comprises $k$ discrete timeslots ($T$), which corresponds to a single training episode. We specify the decision sequence, for example in $T_1$: At the beginning of $T_1$, the frequency network selects a fixed channel based on imperfect sensing. During ${t_1} \in {T_1}$, the power network determines transmission power, which is then used by the modulation network to select the modulation scheme. It proceeds to the next state $t_2$ and makes sequential decisions until $t_l$. After the final short timescale $t_l$, the accumulated state is fed back to update the next long-timescale decision ($T_2$). Since frequency and power/modulation decisions are executed on different timescales, and to reduce the exploration complexity, three decoupled networks are designed to implement the above process. Their corresponding state spaces, action spaces, and reward functions are introduced as follows

\textbullet\hspace{0.3em} \textit{Frequency network}

\textit{a) State Space:} To select the least jammed channel within $T$, the state can be expressed as $\tilde s_T^f = \{ {\tilde s_1},{\tilde s_2}, \ldots ,{\tilde s_n}\} $, where  ${\tilde s_n} = \frac{1}{l}\sum\nolimits_{t = 1}^l {\sum\nolimits_{i = 1}^I {{p_{{\rm{j}},{\rm{i}}}}({f_n}) + {{\sigma ^2}}} } $, and ${{p_{{\rm{j,i}}}}({f_n})}$ denotes the $i$-th jammer's power perceived in the $n$-th channel. 

\textit{b) Action Space:} Select $a_T^f$ from the set $\{ {f_1},{f_2},...,{f_n}\}$. The chosen action remains fixed within $T$.

\textit{c) Reward Function:} The reward should be correlated to the objective. According to (5a), we define the reward function as shown in \eqref{rewardff}, which represents the total throughput within $T$. 
\begin{equation}
	\small
	\label{rewardff}
	\begin{array}{l}
		r_T^f(\tilde s_T^f,a_T^f) = \\
		\sum\limits_{t = 1}^l {\mu  \cdot \psi (a_t^v,\frac{{a_t^p \cdot g_{tr,t}^{{f_n}}}}{{\sum\nolimits_{i = 1}^I {{p_{j,i}}}  + {{\sigma ^2}}}})}  \cdot B \cdot {\log _2}(1 + \frac{{a_t^p \cdot g_{tr,t}^{{f_n}}}}{{\sum\nolimits_{i = 1}^I {{p_{j,i}}}  + {{\sigma ^2}}}})
	\end{array},
\end{equation}
where ${a_t^p}$ and ${a_t^v}$ are the outputs of the power and modulation networks respectively, which will be mentioned later.

\textbullet\hspace{0.3em} \textit{Power network}

\textit{a) State Space:} Given the frequency scheme at $T$, the power network selects its policy at each $t\in T$ based on the chosen channel state. To mitigate the non-stationarity caused by observation aliasing, the state space is augmented with a temporal index ${t_{index}}$, as similar states at different timesteps may require distinct actions due to the adaptive behavior of reactive jammers. For instance, decisions made later in $t$ may adopt higher power to maximize throughput, while earlier ones may use lower power to avoid detection by reactive jammers. Without explicit temporal identifiers, samples corresponding to distinct temporal contexts could be mixed in the replay buffer, causing action ambiguity and unstable training. Therefore, the state space is defined as $\tilde s_t^p = \{ {t_{index}},\sum\nolimits_{i = 1}^I {{p_{{\rm{j}},{\rm{i}}}}(a_T^f) + {{\sigma ^2}}} \}$.

\textit{b) Action Space:} Select $a_t^p$ from the set $\{ {p_1},{p_2},...,{p_u}\}$.

\textit{c) Reward Function:} We take the throughput in $t$ as its immediate reward. 
\begin{equation}
	\label{rewardpp}
	\begin{array}{l}
		r_t^p(\tilde s_t^p,a_t^p) = \\
		\mu  \cdot \psi (a_t^v,\frac{{a_t^p \cdot g_{tr,t}^{{f_n}}}}{{\sum\nolimits_{i = 1}^I {{p_{j,i}}}  + {{\sigma ^2}}}}) \cdot B \cdot {\log _2}(1 + \frac{{a_t^p \cdot g_{tr,t}^{{f_n}}}}{{\sum\nolimits_{i = 1}^I {{p_{j,i}}}  + {{\sigma ^2}}}})
	\end{array}.
\end{equation}

\textbullet\hspace{0.3em} \textit{Modulation network}

\textit{a) State Space:} Since the optimal modulation order depends on instantaneous SJNR, integrating $a_t^p$ with adaptive modulation is essential. Thus, $\tilde s_t^v = \{ {t_{index}},\sum\nolimits_{i = 1}^I {{p_{{\rm{j}},{\rm{i}}}}(a_T^f) + {\sigma ^2}} ,a_t^p\} $.

\textit{b) Action Space:} Select $a_t^v$ from the set $\{ {v_1},{v_2},...,{v_z}\}$.

\textit{c) Reward Function:} Employing the throughput of $t$ as the reward exhibits sparsity, and may induce non-stationarity in the decision process. i.e., low throughput may result from sub-optimal power; even when an appropriate modulation scheme is selected under this condition, the modulation network still receives negative feedback. Guided by the principle of reward shaping \cite{ref31}, and assume that the action space is $\{{v_1},{v_2},{v_3},{v_4}\} $, where the modulation orders are arranged from largest to smallest. Consequently, we reformulate the reward function as follows
\begin{equation}
	\label{rewardv}
	\begin{array}{l}
		r_t^v(\tilde s_t^v,a_t^v) = \\
		\left\{ \begin{array}{l}
			2000 \cdot \lambda  \cdot \Omega (a_t^v),\;\;if \; ({\beta _t})' > {\eta _1},\forall a_t^v\\
			1000 \cdot \lambda  \cdot \Omega (a_t^v),\;\;if \; {\eta _1} > ({\beta _t})' > {\eta _2},a_t^v \in \{ {v_2},{v_3},{v_4}\} \\
			500 \cdot \lambda  \cdot \Omega (a_t^v),\quad if \; {\eta _2} > ({\beta _t})' > {\eta _3},a_t^v \in \{ {v_3},{v_4}\} \\
			200,\quad \quad \quad \quad \quad \; \; if \; ({\beta _t})' < {\eta _3}, a_t^v = {v_4}\\
			0,\quad \quad \quad \quad \quad \quad \; \; otherwise
		\end{array} \right.
	\end{array}.
\end{equation} 

where $({\beta _t})' = 10 \cdot {\log _{10}}({\beta _t})$ denotes the logarithmic representation of ${\beta_t}$ in decibels (dB), and ${\eta _1},{\eta _2},{\eta _3}$ denote the demodulation thresholds, respectively. $\Omega (a_t^v) = \frac{{{{\log }_2}[Or(a_t^v)]}}{{{{\log }_2}[Or(v_{\max }^{valid})]}}$ is the reward factor, where ${Or( \cdot )}$ means the modulation order, and ${v_{\max }^{valid}}$ represents the highest order supported by the current SJNR conditions, $\lambda $ is the sub-optimal penalty factor. The proposed reward function maintains a mapping relationship with the maximization of instantaneous throughput at $t$, thereby achieving decoupling from power decisions. Notably, the modulation scheme is generated through a neural network rather than determined by instantaneous SJNR, as deterministic modulation–SJNR mappings become unreliable under state uncertainty, thereby degrading system performance.

\section{Proposed Projected Gradient Descent-assisted DDQN Algorithm}
In this section, we address the problem where bounded errors in sensing devices lead to uncertain state observations, causing deviations in output actions and even low-reward behaviors. We mathematically reformulate the Bellman operator and prove the existence of a worst-case perturbed state ${\tilde s^*}$. Building on this, we establish a two-step optimization process: (1) Find the worst-case perturbation under state uncertainty and (2) 
Employ regularization constraints with optimal actions in true states as supervisory labels to enhance the neural network’s robustness.

\subsection{Preliminaries and Background}
\textbullet\hspace{0.3em} \textit{Double Deep Q-Network (DDQN):} Existing DRL methodologies include policy gradient-based approaches and value function-based methods. Among value function-based techniques, the DDQN stands as a typical algorithm, demonstrating prominent advantages in handling discrete action space tasks \cite{ref32}. It approximates the action-value $Q(s,a) = r + \gamma {\mathbb{E}_\pi }[Q(s',a')]$ through a neural network, which formally characterizes the cumulative reward of executing action $a$, in state $s$ under policy $\pi $. The $\varepsilon$ -greedy strategy is then used to choose the action with the optimal Q-value, corresponding to the Bellman optimality equation. Moreover, it employs two DNNs to decouple action selection and value estimation, thereby addressing the overestimation bias issue in deep Q-networks (DQN). Specifically, one is the current Q-network, updated with $\theta$, while the other is the target network, updated with $\theta '$. The target value $y$ is formulated by
\begin{equation}
	\label{qvalue}
	y = r + \gamma  \cdot {Q_{{\rm{tar}}}}(s',\mathop {\arg \max }\limits_{a'} {Q_{{\rm{cur}}}}(s',a';\theta );\theta ').
\end{equation} 

Then the networks are trained by minimizing the loss, as shown in \eqref{loss}, where $\mathcal{B}$ is the replay buffer.
\begin{equation}
	\label{loss}
	\mathcal{L}(\theta ) = {\mathbb{E}_{(s,a,s',r) \sim \mathcal{B}}}[{(y - {Q_{{\rm{cur}}}}(s,a;\theta ))^2}].
\end{equation} 

\textbullet\hspace{0.3em} \textit{Projected Gradient Descent (PGD):} PGD is considered one of the strongest white-box attack frameworks in machine learning \cite{ref33}. It searches for the strongest adversarial examples $x'$ within the permitted perturbation range of the input space (e.g., ${\ell _2 }$-norm bounds) to maximize model prediction errors. 
\begin{equation}
	\label{worsts}
	x' = \mathop {\arg \max }\limits_{x' \in {B_{\varepsilon'} }(x)} \mathcal{L}(x',y).
\end{equation}

This approach extends the single-step attack of fast gradient sign method (FGSM) into a multi-step iterative gradient optimization process, and implemented through the introduction of projection operations.
\begin{equation}
	\label{updates}
	{x_{t + 1}} = {\prod _{{B_{\varepsilon'} }(x)}}({x_t} + \alpha  \cdot sign({\nabla _x}\mathcal{L}({x_t},y))).
\end{equation}

where $\alpha$ denotes the single-step perturbation step size, $sign({\nabla _x}\mathcal{L})$ is the sign function along the gradient direction of the loss function to maximize the loss value, and ${\prod _{{B_{\varepsilon'} }}}$ projects the perturbation into the ${\ell _2 }$-norm ball constrained within a radius $\varepsilon'$ centered at the original input $x_0$.

\subsection{Improved Bellman Operator for Uncertain States}
Under perturbed states $\tilde s$, identical policy $\pi$ may select different actions. To enhance policy robustness against state uncertainty, the primary step involves evaluating the policy's value under bounded state perturbations. In this paper, the power and modulation networks receive perturbed states corresponding to the selected channel, while the frequency network processes perturbed states from all channels. For analytical clarity, we concentrate on the frequency network, with analogous reasoning applicable to other networks. We mathematically reformulate the Bellman equations, the perturbed state value function is expressed as
\begin{equation}
	\small
	\label{statevalue}
	\begin{array}{l}
		\tilde V_{\tilde s}^{{\pi _f}}(s) = {\mathbb{E}_{{\pi _f}}}[\sum\limits_{k = 0}^\infty  {{\gamma ^k} \cdot r_{T + k + 1}^f(s_T^f,a_T^f)|s_T^f = s} ]\\
		= \sum\limits_{a \in {A_f}} {{\pi _f}(a|\tilde s)} \sum\limits_{s' \in {S_f}} {p(s'|s,a)}  \cdot [r(s,a) + \gamma \tilde V_{\tilde s}^{{\pi _f}}(s')]
	\end{array},
\end{equation}

where $s$ is the true state, $s'$ is the next state, and $a$ is taken by ${\pi _f}(a|\tilde s)$. In the proposed model, the worst-case scenario is challenging to explicitly define. For example, the agent may refrain from selecting a channel when the sensed jamming power appears high, even though the true state corresponds to a low-power condition. Conversely, it may access the channel when the sensed jamming power is low, while the actual state is subject to strong interference. In fact, under bounded estimation error, the performance degradation admits an upper bound. According to \eqref{statevalue}, the worst-case condition corresponds to minimizing the cumulative reward by leveraging state uncertainty under a given fixed policy $\pi _f$ and state $s$. Therefore, we reformulate the problem as searching for the strongest perturbed state ${\tilde s^*}$,  where $\tilde V_{{{\tilde s}^*}}^{{\pi _f}}(s) = \mathop {\min }\limits_{\tilde s} \tilde V_{\tilde s}^{{\pi _f}}(s)$. Then we propose a worst-case Bellman operator, as shown in \eqref{operator}, which contracts to ${\tilde s^*}$ via Bellman contraction under the fixed policy $\pi _f$, as demonstrated in \textit{Lemma 1}.

\begin{equation}
	\small
	\label{operator}
	\mathcal{T} \tilde V_{\tilde s}^{{\pi _f}}(s): = \mathop {\min }\limits_{\tilde s} \sum\limits_{a \in {A_f}} {{\pi _f}(a|\tilde s)} \sum\limits_{s' \in {S_f}} {p(s'|s,a)}  \cdot [r(s,a) + \gamma \tilde V_{\tilde s}^{{\pi _f}} (s')].
\end{equation}

\textit{Lemma 1:} For any given policy ${{\pi _f}(a|\tilde s)}$, under bounded perturbed state $\tilde s \in {B_{\varepsilon'} }(s)$, the sequence generated by the worst-case Bellman operator $\mathcal{T}$ converges to a unique fixed point $\mathcal{T} \tilde V_{{{\tilde s}^*}}^{{\pi _f}} = \tilde V_{{{\tilde s}^*}}^{{\pi _f}}$, at which the state value is minimized, corresponding to the worst-case ${\tilde s^*}$.

The proof is relegated to Appendix B.

In summary, each policy has a corresponding worst-case ${\tilde s^*}$, our objective reduces to finding an optimal policy ${\pi ^*}$ under the given state $s$, such that its state value in the worst case is higher than that of any others.
\begin{equation}
	\label{optimalpai}
	\tilde V_{{{\tilde s}^*}}^{{\pi ^*}}(s) \ge \tilde V_{{{\tilde s}^*}}^\pi (s),\forall s \in S,\forall \pi.
\end{equation}

Note that the preceding formulation holds for all networks in this paper, encompassing both power and modulation network.

\subsection{Design of the PGD-DDQN Algorithm}
\begin{figure*}[htbp]
	\setlength{\abovecaptionskip}{0.1cm}   
	\centering
	\includegraphics[width=6.8in]{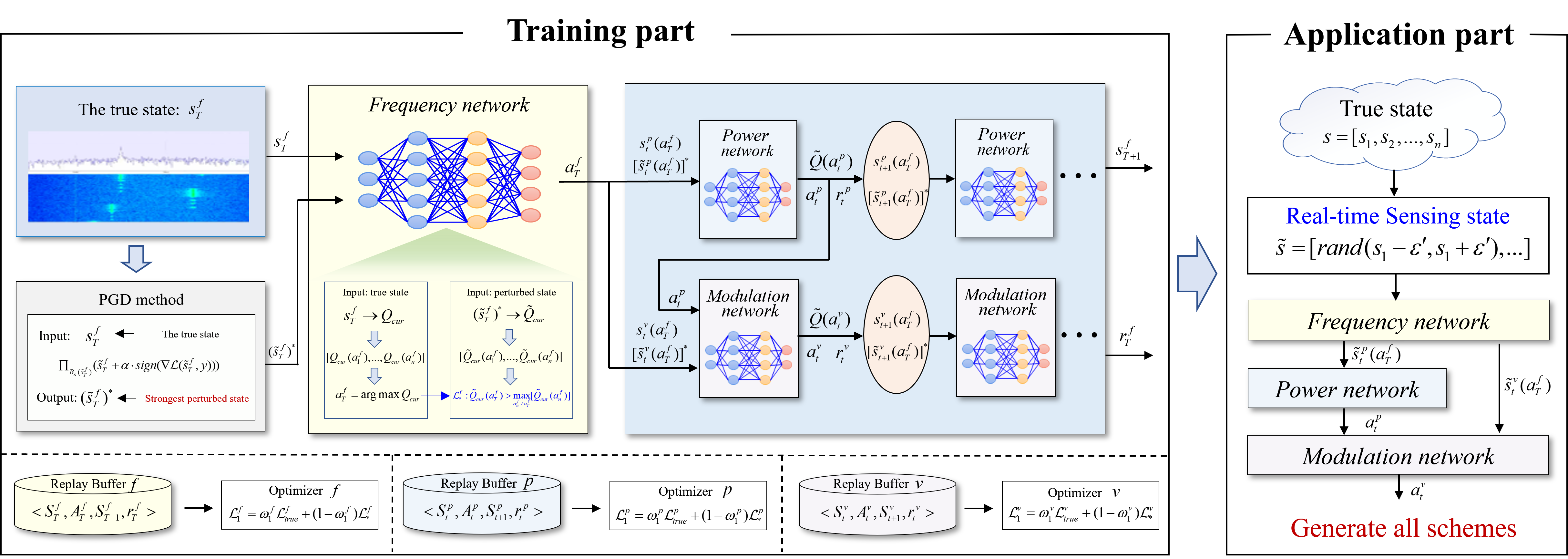}
	\caption{The PGD-DDQN framework. Omitting the target network for enhanced clarity in action decisions. }
	\label{method1}
\end{figure*}

Based on the above discussion, we propose the PGD-DDQN algorithm, which iteratively learns the strongest perturbation through training at each state and performs robust decision-making under such conditions. The entire process operates in two stages, as illustrated in Fig. \ref{method1}.

In the first stage, we construct the strongest perturbed state ${\tilde S}^*$ through the PGD method, grounded in the true environmental state $S$. In the dynamic scenario, the real state of each channel at $t$ be represented as $S = [{s_1},{s_2},...,{s_n}]$. Subject to bounded errors inherent in the sensing device, the actual observed state is modeled as
\begin{equation}
	\label{distorted}
	\tilde S = [rand({s_1} - \varepsilon ' ,{s_1} + \varepsilon ' ),...,rand({s_n} - \varepsilon ' ,{s_n} + \varepsilon ' )].
\end{equation}

where $\varepsilon' = I \cdot \varepsilon$ denotes the error radius for each channel. $rand({s_n} - \varepsilon' ,{s_n} + \varepsilon' )$ denotes the stochastic power perturbation value generated within the channel $n$, mimicking the uncertain state observations induced by imperfect sensing in practical environments.

The PGD method can generate the strongest perturbation under the aforementioned state conditions, and its workflow is illustrated through the frequency network. The true state is $s_T^f = \{ {s_1},{s_2},...,{s_n}\} $ and the corresponding perturbed state is $\tilde s_T^f = \{ {\tilde s_1},{\tilde s_2}, \ldots ,{\tilde s_n}\} $. When feeding $s_T^f$ into the current network, it outputs the ${Q_{cur}}$ value of each action quantifying action efficacy, and we expect to output the optimal action ${(a_T^f)^*} = \arg {\max _a}{Q_{cur}}(s_T^f,a_T^f)$. Under frozen network parameters, we initialize perturbed state $\tilde s_T^f$ as network input, obtaining perturbed ${\tilde Q_{cur}}$ values. The frequency decision task's loss function is formulated as
\begin{equation}
	\label{lossPGD}
	{\cal L}(s_T^f,y) = \mathop {\max }\limits_{a_T^f \ne {{(a_T^f)}^*}} ({\tilde Q_{cur}}(\tilde s_T^f,a_T^f) - y),
\end{equation}

where $y = {\tilde Q_{cur}}(\tilde s_T^f,{(a_T^f)^*})$. Through iterative updates, PGD seeks the strongest perturbed state ${(\tilde s_T^f)^*}$ that maximizes \eqref{lossPGD}, thereby degrading the effectiveness of the original policy, as shown in \eqref{strongeststate}.
\begin{equation}
	\label{strongeststate}
	\tilde s_T^f \leftarrow  {\prod _{{B_{\varepsilon'} }(s_T^f)}}(\tilde s_T^f + \alpha  \cdot sign(\nabla {\cal L}(\tilde s_T^f,y))).
\end{equation}

In the second stage, the goal is to obtain the optimal policy $\pi _f^*$ in ${(\tilde s_T^f)^*}$ that satisfies \eqref{optimalpai}. For DDQN, assume $\varepsilon $-greedy is not considered, the policy operates by selecting the action that maximizes the Q-value output, as shown in \eqref{policy}.
\begin{equation}
	\label{policy}
	{\pi _f}(a|s) = \left\{ \begin{array}{l}
		1,\quad if \; a = \arg {\max _a}{Q_{cur}}(s,a)\\
		0,\quad otherwise
	\end{array} \right..
\end{equation}

During the training phase, the agent can access the true state $s_T^f$ and derive the optimal action $(a_T^f)^*$, which represents the theoretical performance upper bound of the policy. Robustness is achieved if, under the ${(\tilde s_T^f)^*}$, the action output by the neural network aligns with the $(a_T^f)^*$. Building on this principle, we define ${(a_T^f)^*}$ as supervisory label and introduce a regularization term, expressed as
\begin{equation}
	\small
	\label{regulationF}
	{\cal L}_*^f = \max \{ \mathop {\max }\limits_{a_T^f \ne {{(a_T^f)}^*}} ({\tilde Q_{cur}}({(\tilde s_T^f)^*},a_T^f)) - {\tilde Q_{cur}}({(\tilde s_T^f)^*},{(a_T^f)^*}),{\delta}\}, 
\end{equation} 

where $\delta <0$. This regularization term constrains the ${\tilde Q_{cur}}$ values of other actions to ensure the neural network selects ${(a_T^f)^*}$. Besides, according to \eqref{loss}, the canonical loss function under the true state is defined as
\begin{equation}
	\label{lossF}
	\begin{array}{l}
		{\cal L}_{true}^f = [r_T^f + \gamma  \cdot {Q_{{\rm{tar}}}}(s_{T + 1}^f,\\
		\mathop {\arg \max }\limits_{a_{T + 1}^f} {Q_{{\rm{cur}}}}(s_{T + 1}^f,a_{T + 1}^f;\theta );\theta ') - {Q_{{\rm{cur}}}}(s_T^f,a_T^f;\theta ){]^2}
	\end{array}, 
\end{equation} 

Thus, the training loss of the frequency network is shown in \eqref{alllossF}, and $\omega _1^f \in [0,1]$ serves as a robustness trade-off factor, which balances the anti-jamming performance and robustness.
\begin{equation}
	\small
	\label{alllossF}
	{{\cal L}_1^f} = \omega _1^f \cdot L_{true}^f + (1 - \omega _1^f) \cdot L_*^f,
\end{equation} 

Similarly, the loss function of the power and modulation network are defined as 
\begin{equation}
	\small
	\label{alllossP}
	{{\cal L}_1^p} = \omega _1^p \cdot L_{true}^p + (1 - \omega _1^p) \cdot L_*^p,
\end{equation} 
\begin{equation}
	\small
	\label{alllossV}
	{{\cal L}_1^v} = \omega _1^v \cdot L_{true}^v + (1 - \omega _1^v) \cdot L_*^v.
\end{equation} 

Since $\tilde s$ perturbs only action selection without changing the true state, the quadruples are still stored in replay buffers. During training, each network samples a batch from its buffer and employs the PGD method to generate the strongest perturbed states for policy optimization. The training process is summarized in Algorithm 1. After convergence, the pre-trained networks enable real-time robust decisions in uncertain states. The online execution process is illustrated in Fig. \ref{method1}.

\begin{algorithm}[h]
	\caption{The training process of the PGD-DDQN.}
	\begin{algorithmic}[1]
		\STATE Initialize electromagnetic environment\
		\STATE Initialize current Q network for frequency, power, and modulation with parameters ${\theta _f}$, ${\theta _p}$, ${\theta _v}$\	
		\FOR{$episode=1$ to $J_{ep}$}
		\STATE Select an available channel $a_T^f$\
		\FOR{$T=1$ to $k$}
		\FOR{$t=1$ to $l$}
		\STATE Get true state $s_t^p$ and generate ${(\tilde s_t^p)^*}$ by PGD\
		\STATE Select $a_t^p$ via $\varepsilon $-greedy
		\STATE Generate the ${(\tilde s_t^v)^*}$ and obtain $s_t^v$ based on $s_t^v$, $a_t^p$\
		\STATE Select $a_t^v$ via $\varepsilon $-greedy
		\STATE Obtain reward $r_t^p$, $r_t^v$, and the next state  $s_{t + 1}^p$, $s_{t + 1}^v$\
		\STATE Store transition $ < s_t^p,a_t^p,s_{t + 1}^p,r_t^p >$ in the replay buffer ${\mathcal{B}_\mathrm{p}}$ and $ < s_t^v,a_t^v,s_{t + 1}^v,r_t^v >$ in ${{\mathcal{B}}_\mathrm{v}}$\
		\STATE Sample minibatches from ${{\mathcal{B}}_\mathrm{p}}$, ${{\mathcal{B}}_\mathrm{v}}$ and update the power and modulation network by \eqref{alllossP}, \eqref{alllossV}\
		\ENDFOR
		\STATE Calculate the state $s_T^f$ and generate the ${(\tilde s_T^f)^*}$\
		\STATE Select $a_T^f$, obtain $r_T^f$ by \eqref{rewardff}, and next state $s_{T + 1}^f$\
		\STATE Store transition $ < s_T^f,a_T^f,s_{T + 1}^f,r_T^f >$ in ${{\mathcal{B}}_\mathrm{f}}$\
		\STATE Sample minibatch from ${{\mathcal{B}}_\mathrm{f}}$ and update the frequency current network by \eqref{alllossF}\
		\ENDFOR
		\STATE Update the target Q network every certain episodes
		\ENDFOR
		\label{PGDDDQN}
	\end{algorithmic}
\end{algorithm}

\section{Robustness Enhancement for DDQN via Nonlinear Q-Compression}
In the previous section, the PGD-DDQN algorithm theoretically ensured the lower bound of system performance. In this section, we leverage the property of DNNs where bounded inputs lead to bounded outputs. Specifically, we embed a nonlinear compression function in the network's output layer to constrain the upper and lower bounds of Q-values across different actions. Building on this, a Q-Separation Regularization (QSR) term is introduced to keep the Q-value intervals of the optimal action clearly separated from those of the other actions. This approach fundamentally establishes immunity to bounded estimation errors in the sensing device.

\subsection{Background on Interval Bound Propagation}
Interval Bound Propagation (IBP) is a formal verification method based on interval arithmetic \cite{ref34}. By modeling the input perturbations and forward-propagating the intervals layer by layer to calculate the deterministic upper and lower bounds of the network output. Thus it is possible to explicitly optimize the robustness of neural networks in the training phase.

Given an n-dimensional state vector $\mathbf{S} = [{s_1},{s_2},...,{s_n}]$, each dimension is subject to bounded perturbations. Accordingly, the perturbed input can be represented as an interval tensor ${\mathbf{S}_{\varepsilon'} } = [({\underline{s}_1},{\bar s_1}),({\underline{s}_2},{\bar s_2}),...,({\underline{s}_n},{\bar s_n})]$, where $({\underline{s}_n},{\bar s_n}) = [{s_n} - \varepsilon' ,{s_n} + \varepsilon' ]$. Through propagation across linear transformation layers and nonlinear activation functions within the feedforward neural network, deterministic bounds for the y-dimensional action Q-values ${Q_{\varepsilon'} } = [({\underline{Q}_1},{\bar Q_1}),({\underline{Q}_2},{\bar Q_2}),...,({\underline{Q}_y},{\bar Q_y})]$ are derived at the output layer. The proof process is presented in Appendix C.

\subsection{Design of the NQC-DDQN algorithm}
\begin{figure*}[htbp]
	\setlength{\abovecaptionskip}{0.1cm}   
	\centering
	\includegraphics[width=6.2in]{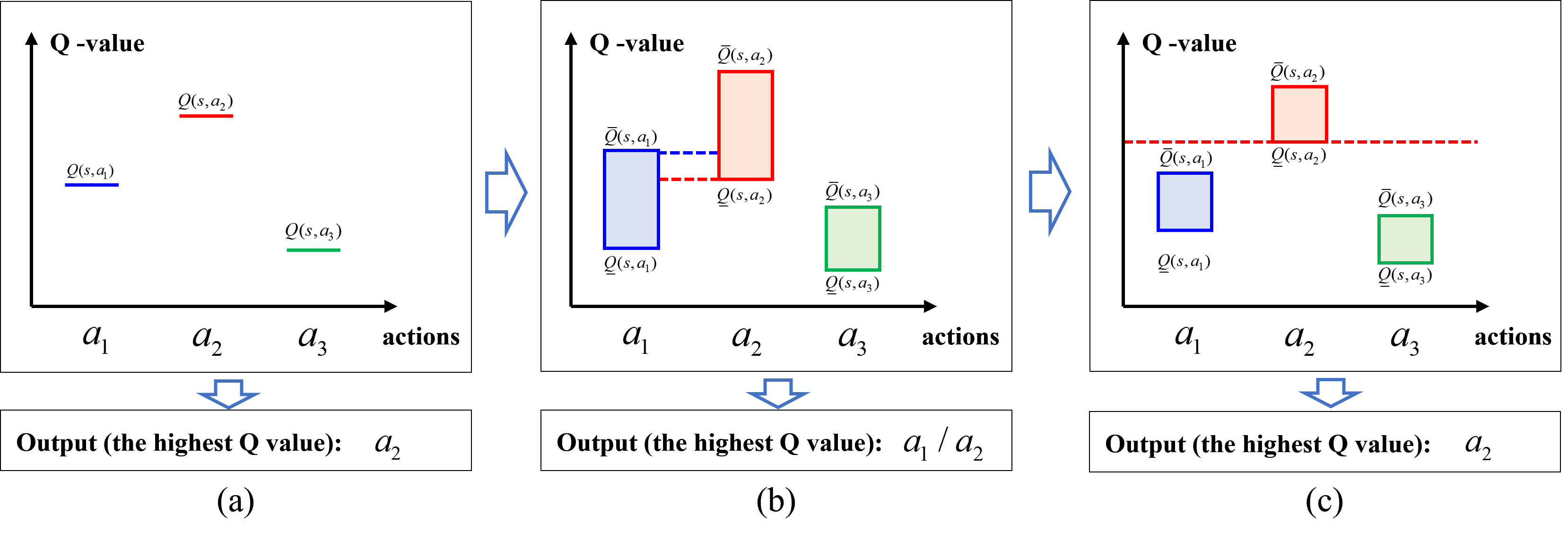}
	\caption{The Q-values of different actions output by the neural network under a given state.}
	\label{method2}
\end{figure*}

This method analyzes the robustness of Q-value estimation in DRL. In deterministic true states, DRL models generate constant Q-value estimates for different actions under a given state, as depicted in Fig. \ref{method2}(a). Nevertheless, when there are bounded errors in the sensing device, the Q-value estimations exhibit various ranges of forms, as shown in Fig. \ref{method2}(b). Notably, according to \eqref{policy}, identical policies may yield divergent action outputs (e.g., $a_1$, $a_2$). If the actual value of $a_1$ is low, there exists a risk of potential decision failure. 

Our contribution lies in establishing a Q-value correction mechanism based on nonlinear compression. By applying nonlinear transformations to the Q-values of different actions across various states, the proposed method enforces that the lower bound of the optimal action’s Q-value exceeds the upper bounds of all other actions, as illustrated in Fig.~\ref{method2}(c). Within the predefined range of sensing errors, this approach guarantees persistent output of optimal actions in true states, thereby enhancing decision robustness. Based on this we propose the NQC-DDQN algorithm.

We still take the frequency network as an example for elaboration, implementing the nonlinear compression process of Q-values from two perspectives. 

First, inspired by the concept of safe reinforcement learning \cite{ref35}, a differentiable function is embedded in the output layer of the neural network, as shown in \eqref{nonlinearfunction}
\begin{equation}
	\label{nonlinearfunction}
	g(Q({f_n});c,\psi ) = c + (Q({f_n}) - c) \cdot {e^{ - \psi  \cdot |Q({f_n} - c)|}},
\end{equation} 
where $Q({f_n}) \in [\underline{Q}({f_n}),\bar Q({f_n})]$ represents the Q-value at different actions, $c$ is the interval center, and $\psi $ is the compression coefficient. This function compresses the Q-value interval toward the center $c$, where values farther from the center experience stronger compression, and the compression magnitude decays exponentially with distance.

\begin{figure}[htbp]
	\centering
	\includegraphics[width=3.5in]{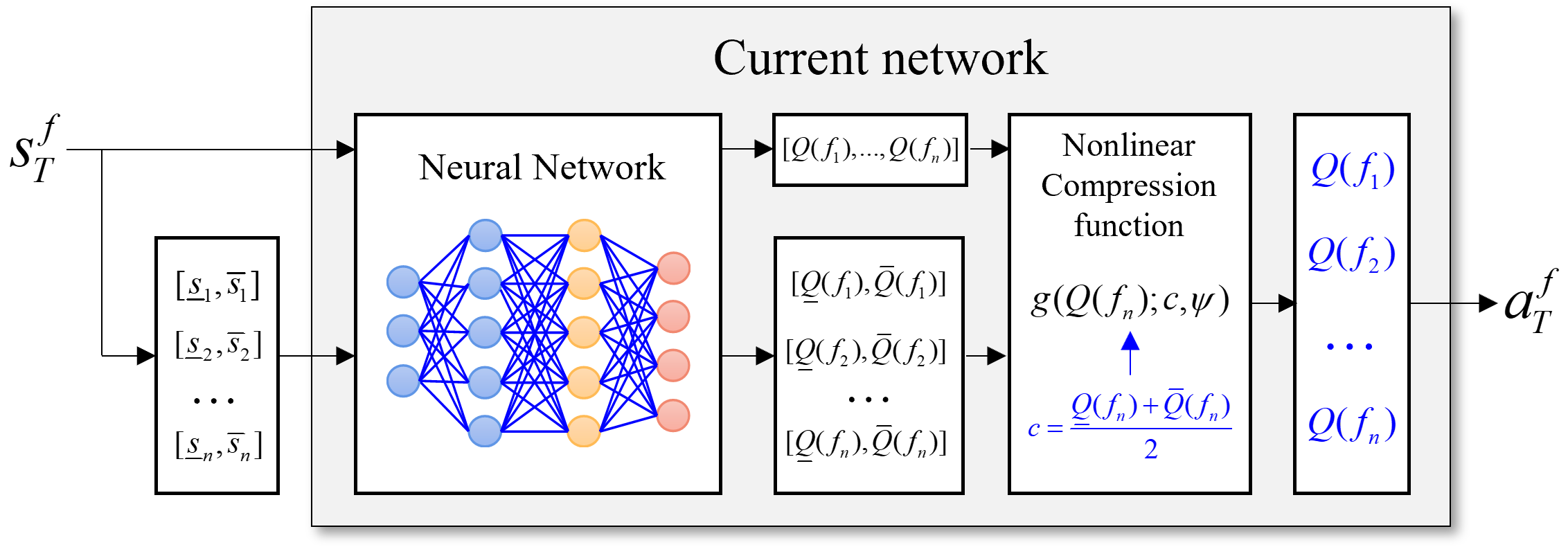}
	\caption{Architecture of the frequency current network.}
	\label{compression}
\end{figure}

Therefore, the current network structure of the frequency network is restructured in Fig. \ref{compression}. where the real deterministic state is input and outputs the Q-values $Q({f_n})$ of each discrete action. Additionally, the upper and lower bounds of state uncertainty are simultaneously fed into the network. By applying the IBP method, the Q-value ranges for each action are derived through forward propagation. Building on this, \eqref{nonlinearfunction} is utilized to perform nonlinear compressive mapping of the Q-value ranges. This function contracts each Q-value interval toward its midpoint, and thereby reduces the overlap degree of Q-value ranges among different actions.

Note that while the aforementioned compressive function can reduce Q-value aliasing, it cannot guarantee complete elimination of interval overlap risks between optimal and other actions. To address this, secondly, we formally define the misleading action set as shown in \eqref{misleadmethod2}. Then we incorporate QSR as a regularization term into the loss function. Through gradient backpropagation mechanism, this enables dynamic adjustment of network parameters to enforce rectification of Q-values between optimal and misleading actions.
\begin{equation}
	\label{misleadmethod2}
	\begin{array}{*{20}{l}}
		{{{\cal A}_{mislead}^f}: = \{ (a_T^f)'|{{\bar Q}_{cur}}(s_T^f,(a_T^f)') > {{\underline{Q}}_{cur}}(s_T^f,{{(a_T^f)}^*})}\\
		{\begin{array}{*{20}{c}}
				{\begin{array}{*{20}{c}}
						{}&{}
				\end{array}}&{}
			\end{array},(a_T^f)' \in {{\cal A}_f},(a_T^f)' \ne {{(a_T^f)}^*}\} }
	\end{array},
\end{equation} 
where ${{{(a_T^f)}^*}}$ denotes the optimal action under true states, serving as the theoretical upper bound of the policy’s performance. While ${(a_T^f)'}$ represents the misleading action set induced by perturbed states, where the upper Q-value bound of these actions exceeds the lower Q-value bound of ${{{(a_T^f)}^*}}$. Thus, the QSR is defined in \eqref{QSR}.
\begin{equation}
	\label{QSR}
	{({{\cal L}^f})^\prime } = \sum\limits_{{{(a_T^f)}^\prime }} {\max [{{\bar Q}_{cur}}(s_T^f,{{(a_T^f)}^\prime }) - {{\underline{Q}}_{cur}}(s_T^f,{{(a_T^f)}^*})]} .
\end{equation} 

This regularization term is designed to ensure ${{\cal A}_{mislead}^f} = \emptyset $, thereby ensuring consistent output of optimal actions under perturbed states to achieve robust decision-making. By integrating \eqref{lossF}, the loss function of the frequency network in the proposed method is formulated as
\begin{equation}
	\label{method2F}
	{\cal L}_2^f = \omega _2^f \cdot {\cal L}_{true}^f + (1 - \omega _2^f) \cdot ({L^f})',
\end{equation} 

The power and modulation networks, operating at short timescale $t$, define misleading action sets ${{\cal A}_{mislead}^p}, {{\cal A}_{mislead}^v}$, and enforce mandatory Q-value corrections, with the loss functions formulated respectively as follows
\begin{equation}
	\label{method2P}
	{\cal L}_2^p = \omega _2^p \cdot {\cal L}_{true}^p + (1 - \omega _2^p) \cdot ({L^p})',
\end{equation} 
\begin{equation}
	\label{method2V}
	{\cal L}_2^v = \omega _2^v \cdot {\cal L}_{true}^v + (1 - \omega _2^v) \cdot ({L^v})'.
\end{equation} 

The training process of the NQC-DDQN algorithm is summarized in Algorithm 2. All networks employ IBP to calculate action value ranges and applies Q-value rectification mechanisms. Specifically, the Q-value range is contracted by rewriting the neural network structure, while a regularization term is introduced to enforce Q-value boundary separation between the optimal and suboptimal actions. Once the training is accomplished, its application part is identical to that of the PGD-DDQN algorithm, as illustrated in Fig. \ref{method1}.
\begin{algorithm}[h]
	\caption{The training process of the NQC-DDQN.}
	\begin{algorithmic}[1]
		\STATE Initialize electromagnetic environment\
		\STATE Initialize current Q network for frequency, power, and modulation with parameters ${\theta _f}$, ${\theta _p}$, ${\theta _v}$\	
		\STATE Rewrite the architecture of current Q networks by Fig. \ref{compression}.
		\FOR{$episode=1$ to $J_{ep}$}
		\STATE Select an available channel $a_T^f$\
		\FOR{$T=1$ to $k$}
		\FOR{$t=1$ to $l$}
		\STATE Get the true state $s_t^p$ and calculate the $[{{\underline{Q}}_{cur}}(s_t^p,a_t^p),{\bar Q_{cur}}(s_t^p,a_t^p)]$ by IBP method\
		\STATE Select $a_t^p$, and ${(a_t^p)^*} := \arg {\max _{a_t^p}}{Q_{cur}}(s_t^p,a_t^p)$ 
		\STATE Confirm the misleading actions of the set ${{\cal A}_{mislead}^p}$
		\STATE Generate the state $s_t^v$ based on $s_t^v$, $a_t^p$, and calculate the $[{{\underline{Q}}_{cur}}(s_t^v,a_t^v),{\bar Q_{cur}}(s_t^v,a_t^v)]$ by IBP method\
		\STATE Select $a_t^v$, and ${(a_t^v)^*} := \arg {\max _{a_t^v}}{Q_{cur}}(s_t^v,a_t^v)$ 
		\STATE Confirm the set ${{\cal A}_{mislead}^v}$
		\STATE Obtain reward $r_t^p$, $r_t^v$, and the next state  $s_{t + 1}^p$, $s_{t + 1}^v$\
		\STATE Store transition $ < s_t^p,a_t^p,s_{t + 1}^p,r_t^p >$ in the replay buffer ${\mathcal{B}_\mathrm{p}}$ and $ < s_t^v,a_t^v,s_{t + 1}^v,r_t^v >$ in ${{\mathcal{B}}_\mathrm{v}}$\
		\STATE Sample minibatches from ${{\mathcal{B}}_\mathrm{p}}$, ${{\mathcal{B}}_\mathrm{v}}$ and update the power and modulation network by \eqref{method2P}, \eqref{method2V}\
		\ENDFOR
		\STATE Get $s_T^f$ and obtain $[{{\underline{Q}}_{cur}}(s_T^f,a_T^f),{\bar Q_{cur}}(s_T^f,a_T^f)]$ \
		\STATE Select $a_T^f$, derive ${{{(a_T^f)}^*}}$ and confirm the set ${{\cal A}_{mislead}^f}$, then obtain reward $r_T^f$ and next state $s_{T + 1}^f$\
		\STATE Store transition $ < s_T^f,a_T^f,s_{T + 1}^f,r_T^f >$ in ${{\mathcal{B}}_\mathrm{f}}$\
		\STATE Sample minibatch from ${{\mathcal{B}}_\mathrm{f}}$ and update the frequency current network by \eqref{method2F}\
		\ENDFOR
		\STATE Update the target Q network every certain episodes
		\ENDFOR
		\label{NQCDDQN}
	\end{algorithmic}
	
\end{algorithm}
\section{Simulation results}
\begin{figure}[htbp]
	\centering
	\includegraphics[width=3.2in]{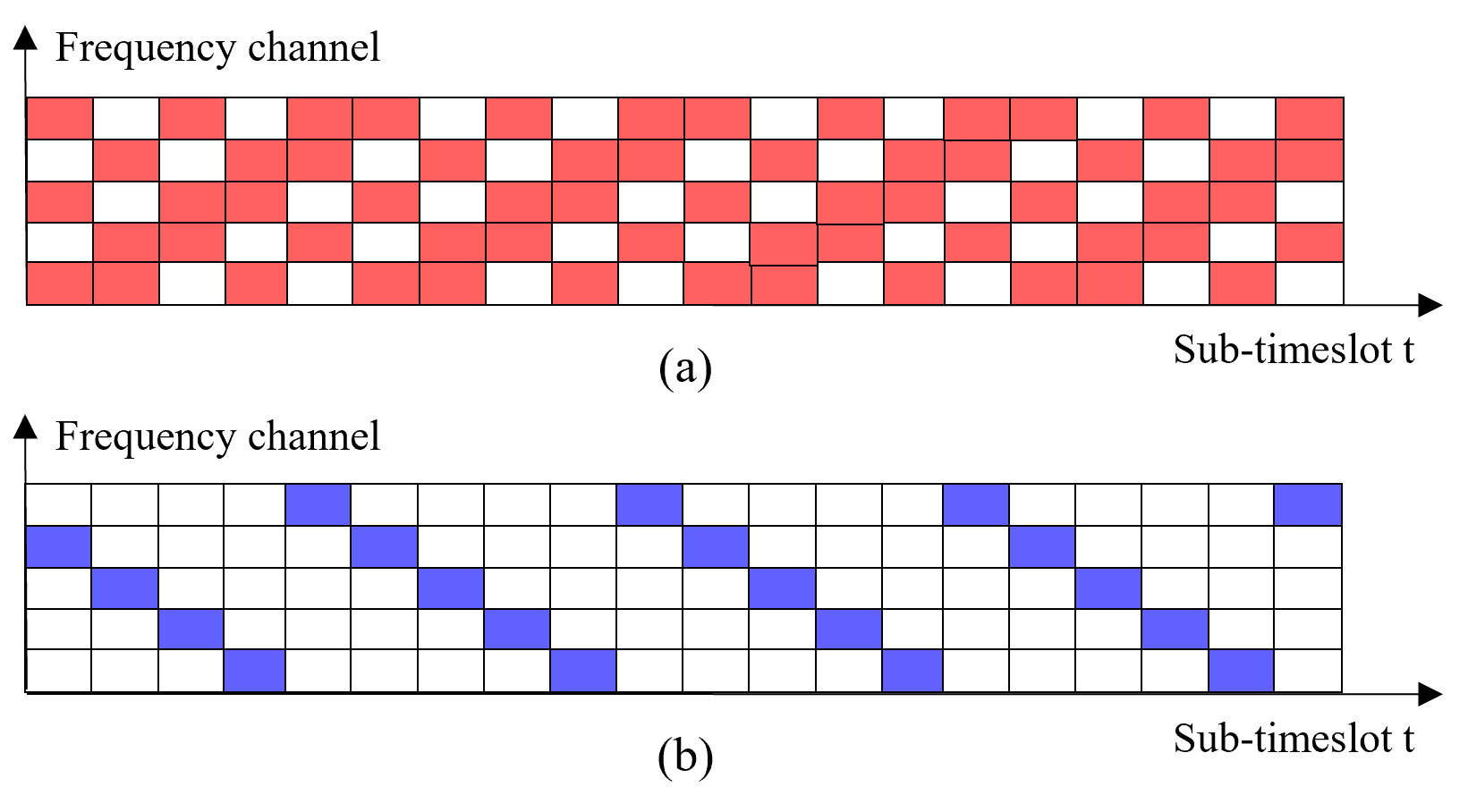}
	\caption{The time-frequency distribution (displaying the first 20 $t$ slots). (a) The jamming pattern of jammer 2. (b) The jamming pattern of jammer 3.}
	\label{jam1and2}
\end{figure}

This section evaluates the robust anti-jamming performance of the proposed algorithms in a single transmitter–receiver link scenario involving three jammers with distinct locations and capabilities. The total communication period is $M=30$ ms, comprising long-timescale $T=3$ ms and short-timescale $t=1$ ms. Jammer 1 employs cognitive narrowband jamming with a detection threshold; it attacks the next slot $t_{l+1}$ if the detected transmit power in $t_{l}$ exceeds the threshold. Jammers 2 and 3 perform periodic comb-sweeping jamming as shown in Fig. \ref{jam1and2}. All decision variables are defined within discrete action spaces. The available spectrum 500-600MHz is divided into 5 orthogonal channels (the bandwidth is 10MHz). The transmit power levels and modulation schemes are discretized based on the power intensity and modulation order, respectively. Note that power parameters are expressed in dBm and can be converted from watts by ${P_{dbm}} = 10 \cdot {\log _{10}}(1000 \cdot {P_{W}})$. Besides, the sensing device deployed at the receiver is subject to inherent bounded measurement errors, and the $\varepsilon = 10$ W. Detailed simulation parameters are provided in Table 1.

\begin{table}[htbp]
	\centering
	\caption{parameter of simulation environment}
	\begin{tabular}{l l}
		\toprule
		Parameters & Value \\
		\midrule
		The whole communication cycle $M$ & 30 ms \\
		Long timescale (timeslot) $T$ \cite{ref36} & 3 ms \\
		Short timescale (sub-timeslot) $t$ & 1 ms \\
		Positions for transmitter and receiver & (0, 5) km, (5, 5) km \\
		Positions for Jammer 1 & (4, 10) km \\
		Positions for Jammer 2,3 & (2, 1.5) km, (9, 8) km \\
		Communication spectrum range & (500, 600) MHz \\
		Bandwidth $B$ & 10 MHz \\
		The required threshold ${\mu _{\mathrm{th}}}$ & 1 Mb/s \\
		Transmit power of transmitter & [25, 30, 35, 40, 45, 50] dBm \\
		Transmit power of Jammer 1,2,3 & [53, 45, 45] dBm \\
		Detection threshold of Jammer 1 & -55 dBm \\
		Modulation $V$ & [BPSK,8PSK,16QAM,64QAM] \\
		Demodulation threshold ${\eta _1}, {\eta _2}, {\eta _3}$ & [15, 10, 5] dB \\
		The error boundary radius ${\varepsilon}$ & 10 W \\
		The sub-optimal penalty factor $\lambda $ & 0.7 \\	
		Environment noise & -80 dBm \\
		\bottomrule
	\end{tabular}
	\label{tab:parameters}
\end{table}

In this paper, the frequency, power, and modulation networks employ a unified hidden-layer architecture comprising three fully connected layers with 32 neurons and ReLU activations, while differing only in the dimensionality of their input and output layers. The hyperparameters are configured as follows: 2,000 training episodes, a learning rate initialized at 0.01 with episode-progressive decay, and the discount factor $\gamma = 0.3$ \cite{ref37}. Notably, action decisions operate at distinct timescales and generate varying amounts of data. Therefore, we design differentiated experience replay buffer capacities: 2,000 samples for the frequency network versus 3,000 samples for the power and modulation networks, the mini-batch size is 128. Besides, after numerous simulation experiments, for the PGD-DDQN, the iteration of PGD is 20, and the single-step perturbation step size $\alpha=1/20$. The ${\delta}=-100$ and $\omega _1^f, \omega _1^p, \omega _1^v$ are set to 0.5. For the NQC-DDQN, the compression coefficient $\psi = 0.005$, and $\omega _2^f, \omega _2^p, \omega _2^v$ are set to 0.5, representing the equilibrium between the anti-jamming performance in true state and the robustness of the schemes.
 
\subsection{Ablation Experiment}
\begin{figure}[htbp]
	\centering
	\includegraphics[width=3.2in]{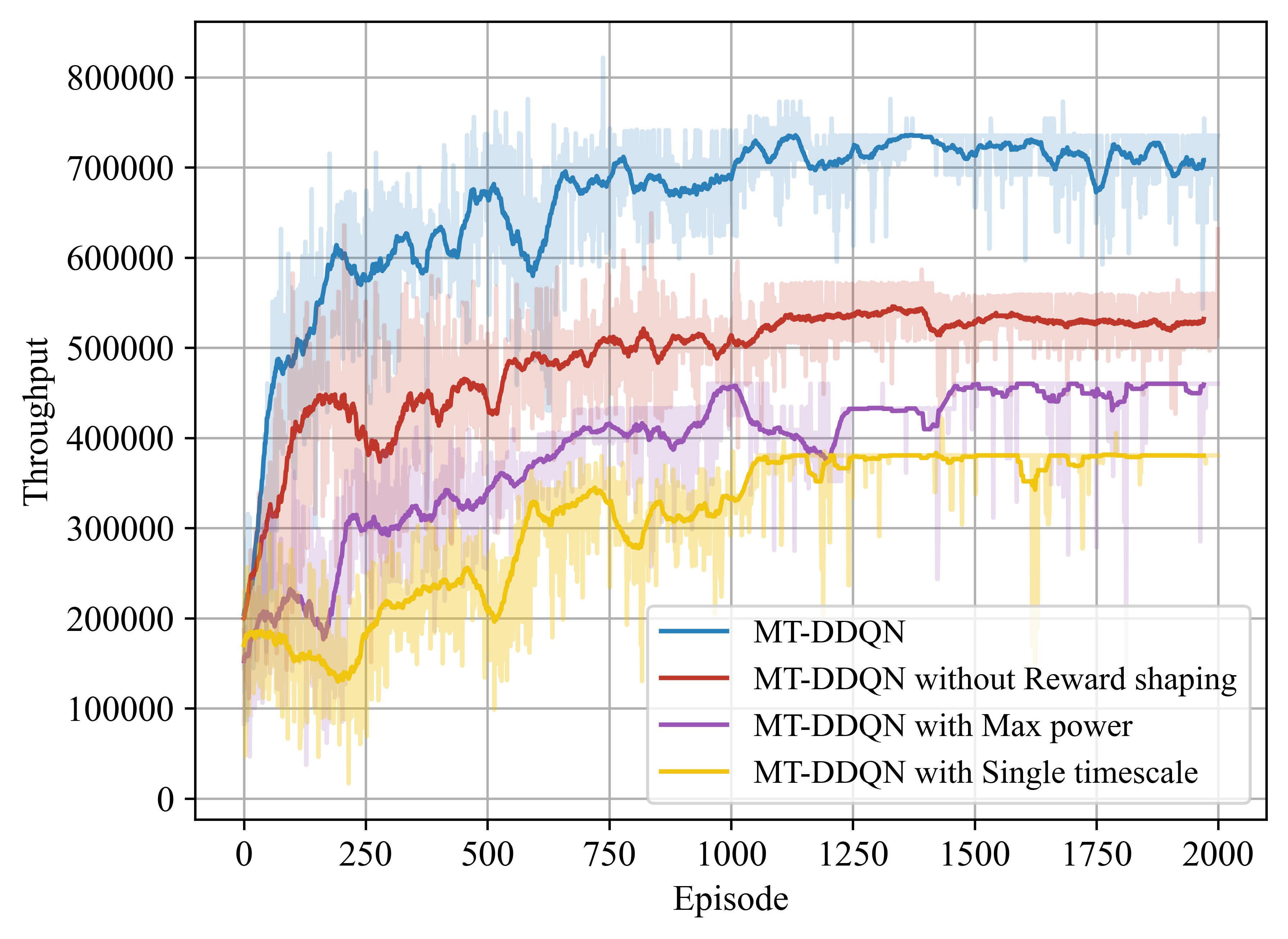}
	\caption{Ablation analysis on key components of the system model.}
	\label{sim1_ablation}
\end{figure}

In this section, we conduct ablation experiments, as shown in Fig. \ref{sim1_ablation}, to analyze the rationality of the key components in the proposed system model. Note that all simulations are performed under true states, and all variants share identical hyperparameters to ensure fairness. The Multi-timescale DDQN (MT-DDQN) algorithm represents the full version of our proposed model. First, we analyze the necessity of multi-timescale decision-making. Since frequency cannot be switched within the short timescale $t$, enforcing a unified timescale causes power and modulation to remain fixed over $T$, thereby limiting adaptability to rapidly varying jamming. Consequently, the MT-DDQN achieves a 93.12\% throughput gain over the single timescale variant. Second, we examine the fixed maximum-power transmission variant (purple line), whose performance is suboptimal. While higher transmit power can improve instantaneous throughput, it simultaneously exposes the transmitter to a higher detection probability and more intensive jamming. Thus under strong jamming conditions, adaptive power control remains essential for maintaining transmission concealment.

Finally, due to the modulation network does not incorporate reward shaping (sharing the same reward as the power network), the upward trend of the training curve is sluggish (red line), which is caused by decision non-stationarity. When the power policy yields low rewards, even superior modulation decisions can produce negative feedback for parameter updates. After introducing reward shaping, the modulation network is optimized under a reward function driven by actual experience, effectively decoupling modulation from power and converging to a superior solution. In conclusion, under rapidly varying and strong jamming conditions, optimization must be performed at different timescales according to the response latency of each action. Moreover, optimizing the power and the reward shaping of the modulation network both contribute significantly to the system's performance.

\subsection{Comparison of Anti-jamming Performance}
To verify the anti-jamming performance of the proposed algorithms, we compared the following methods.

\textbullet\hspace{0.3em} \textit{MT-DDQN:} Train under the true states as the baseline.

\textbullet\hspace{0.3em} \textit{PGD-DDQN:} The proposed in Section \uppercase\expandafter{\romannumeral3}.

\textbullet\hspace{0.3em} \textit{NQC-DDQN:} The proposed in Section \uppercase\expandafter{\romannumeral4}.

\textbullet\hspace{0.3em} \textit{Greedy:} Select the optimal action at the current timestep.

\textbullet\hspace{0.3em} \textit{Random:} Randomly select actions in each timeslot.

\begin{figure}[htbp]
	\centering
	\includegraphics[width=3.2in]{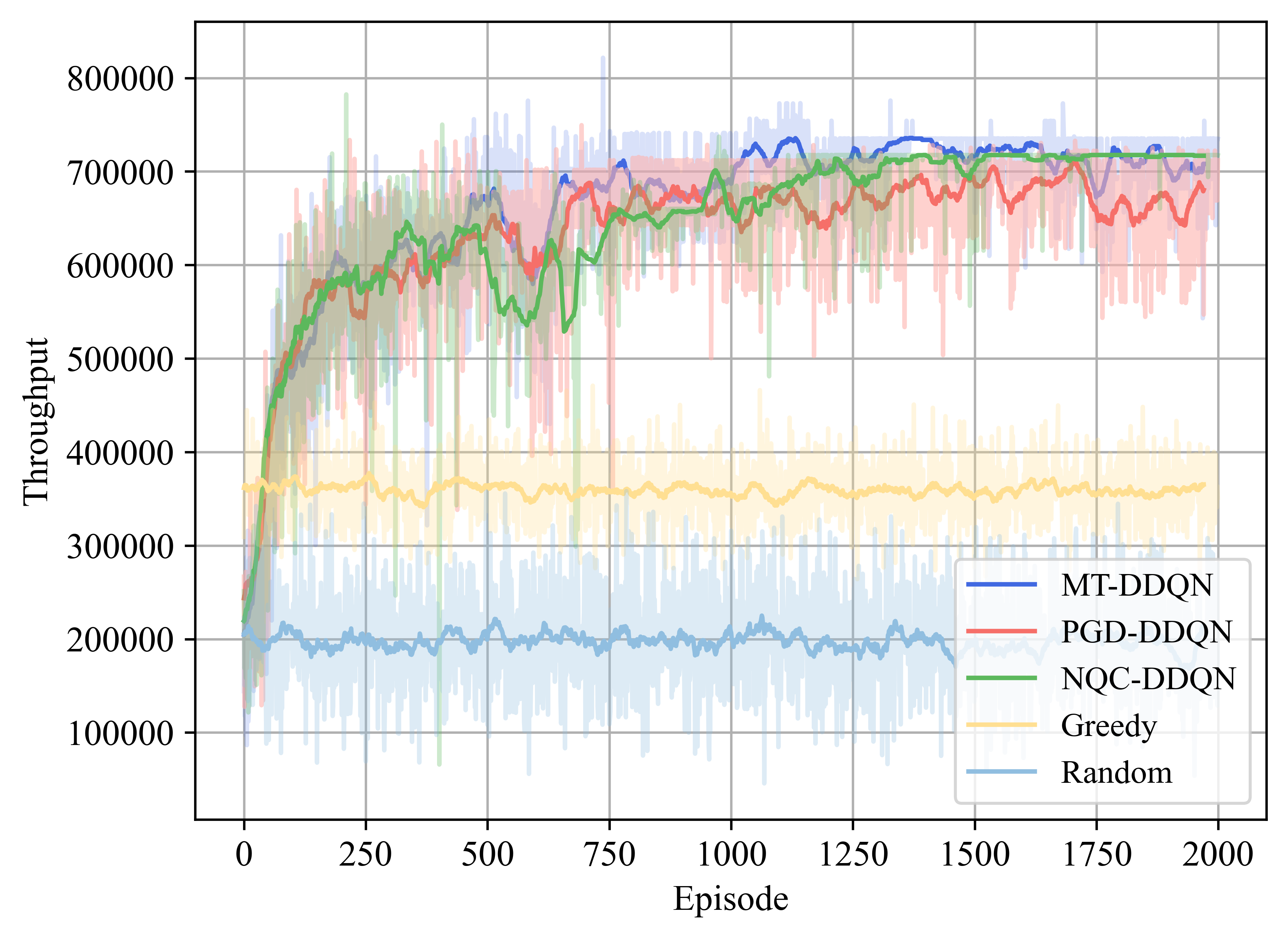}
	\caption{Anti-jamming performance comparison.}
	\label{sim2_compare}
\end{figure}

The simulation outcomes are presented in Fig. \ref{sim2_compare}. The greedy algorithm exhibits performance fluctuations and overall inferior results. This behaviour arises not only from its suboptimal action selection mechanism but, more critically, from its cognitive deficiency in handling state uncertainty. Specifically, the algorithm implicitly assumes perfect sensing information; as a result, its action selection under uncertain states—based on ideal sensing assumptions—often leads to partial decision failures. The inconsistent outcomes observed across repeated decision processes further confirm the limitations of such conventional approaches in uncertain environments.

The MT-DDQN algorithm, trained under ideal state perception conditions, demonstrates the best anti-jamming performance. The performances of the PGD-DDQN and NQC-DDQN algorithms are similar and slightly inferior to that of the MT-DDQN. This is attributed to the introduction of a regularization constraint term during the neural network training phase, which may cause the selection of a suboptimal action with better stability in certain states. Note that although slightly conservative in some states, they incur only 3.11\% and 2.38\% performance losses, respectively, compared with the MT-DDQN, which remain within an acceptable range.

\subsection{Robustness Analysis of the Proposed Algorithms}
\begin{figure}[htbp]
	\centering
	\includegraphics[width=3.5in]{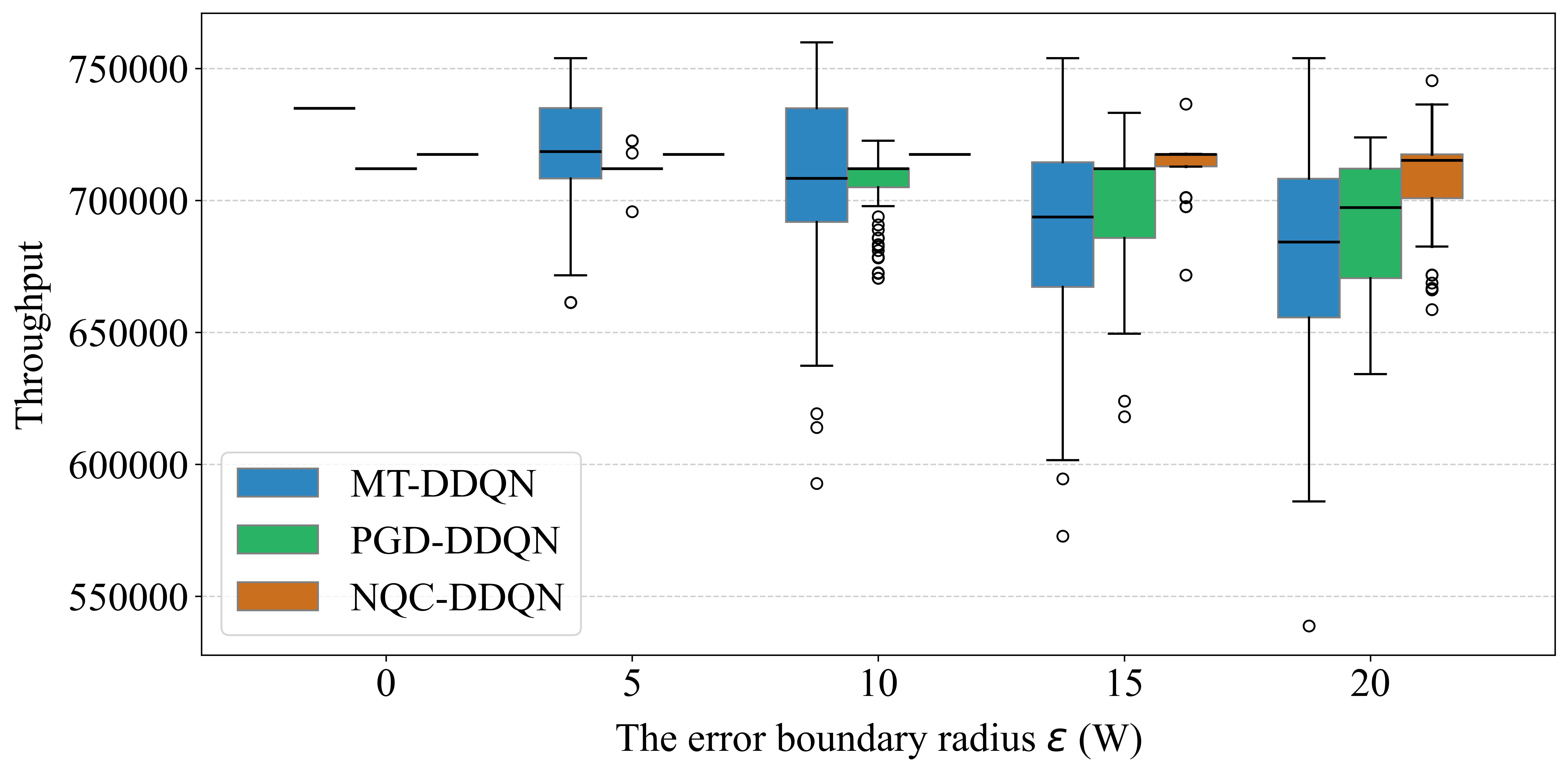}
	\caption{Statistical results under various perturbation levels. Each box represents the distribution of cumulative throughput collected from 200 runs.}
	\label{box}
\end{figure}

This section conducts an analysis of the robustness of the proposed algorithms, which constitutes the core of this paper. We compared the MT-DDQN, PGD-DDQN, and NQC-DDQN algorithms. After convergence during the training stage, the neural network models are deployed to undertake decision-making tasks in actual uncertain environments. As shown in Fig. \ref{box}, the algorithm’s performance is tested over 200 runs under varying perturbation levels (error radius ${\varepsilon}$), where the perception errors are randomly generated within the error radius in each run to simulate state uncertainty.

When ${\varepsilon} = 0$, implying perfect sensing, all algorithms demonstrate stable performance. Within the perturbation escalation regime $0< {\varepsilon} < 10$, the MT-DDQN exhibits clear dynamic instability, and its lowest performance deteriorates significantly, indicating that state uncertainty causes the selection of poor strategies, which is intolerable for reliable system operation. In contrast, the PGD-DDQN algorithm improves the lower bound of performance by rectifying actions under the state corresponding to the strongest perturbation, and it shows relatively minor fluctuations. The NQC-DDQN demonstrates superior robustness by maintaining strict decision invariance within the predefined bounded-error domain. When ${\varepsilon}=10$, both proposed algorithms exhibit higher median throughput and considerably narrower interquartile ranges (IQRs) than the MT-DDQN algorithm, reflecting enhanced robustness and improved overall performance under state uncertainty. When ${\varepsilon}>10$, that is, when the error radius exceeds the predefined range,  while all algorithms exhibit degradation in robustness metrics, the NQC-DDQN shows the least degradation trend and ensure anti-jamming capabilities.

In addition, we observe a phenomenon where state uncertainty occasionally leads to performance exceeding the baseline (${\varepsilon}=0$). This can be attributed to state perturbations altering the agent’s actions, as the converged policy under perfect sensing is not necessarily globally optimal. Therefore, it is possible that state perturbations may lead to better performance. However, when examining the third-quartile (Q3) throughput of the MT-DDQN algorithm, it is observed to decrease with increasing uncertainty and remain below the baseline, indicating that over 75\% of the decisions are adversely affected. Such results underscore the need for more robust methods to ensure stability.

\begin{figure}[htbp]
	\centering
	\includegraphics[width=3.5in]{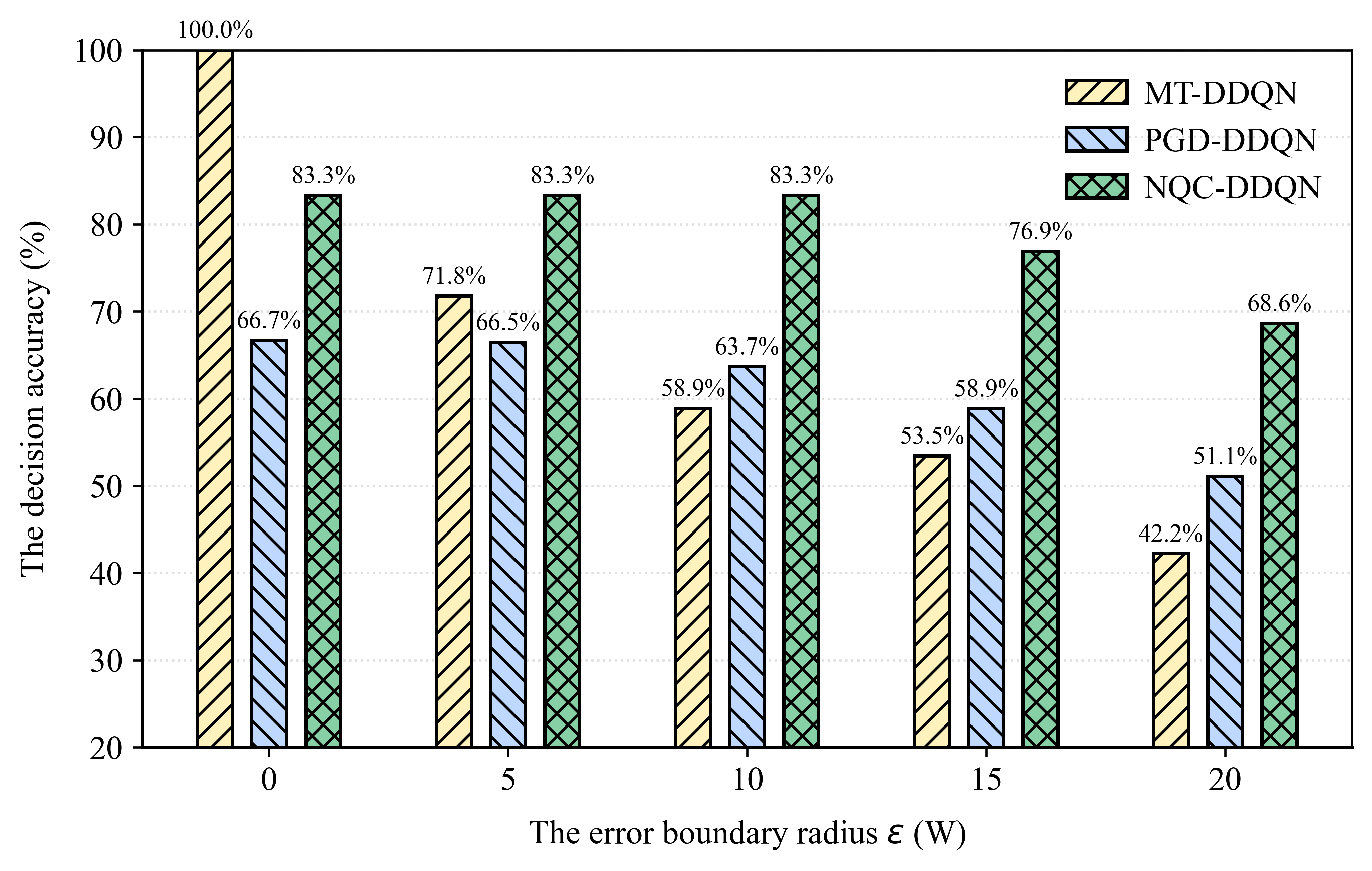}
	\caption{The accuracy of different algorithms under various perturbations.}
	\label{success}
\end{figure}

To intuitively evaluate the robustness of the proposed algorithms, we quantitatively analyse the decision accuracy of PGD-DDQN and NQC-DDQN. The optimal actions $a_{MT}^*$ obtained by MT-DDQN at ${\varepsilon}=0$ serve as the benchmark. Comparisons are performed across frequency, power, and modulation decisions. Specifically, if the tested algorithm selects the same action under the corresponding state, it is regarded as a correct decision; otherwise, it is classified as a decision deviation. The decision accuracy is defined in \eqref{accuracy}.

\begin{equation}
	\small
	\label{accuracy}
	A{\rm{cc}}uracy = \frac{1}{\kappa}\sum\limits_1^{\kappa} {\frac{{\delta ({a_{MT/PGD/NQC}} = a_{MT}^*)}}{{{Lens(a_{MT}^*)}}}}  \times 100\% ,
\end{equation} 

where $a_{MT/PGD/NQC}$ denotes the actions generated by different algorithms across all states in an episode, and $\delta(\cdot)=1$ if true. $\kappa$ represents the number of tests, which is set to 200, and the averaged results are used to evaluate robustness. As shown in Fig.~\ref{success}, the MT-DDQN exhibits the most significant decline in decision accuracy with increasing uncertainty, dropping to 42.2\% at ${\varepsilon}=20$, indicating that more than half of the decisions deviate due to state perturbations. The PGD-DDQN achieves 66.7\% accuracy at ${\varepsilon}=0$, primarily enhancing the lower bound of the performance. It tends to select relatively conservative sub-optimal actions, yet it exhibits good stability. The NQC-DDQN, benefiting from its nonlinear Q-value interval adjustment, maintains high decision accuracy and strong stability, remaining around 83.3\% within the predefined bounded-error domain. Nevertheless, there are still certain decision deviations.

\begin{figure*}[htbp]
	\centering
	\includegraphics[width=7in]{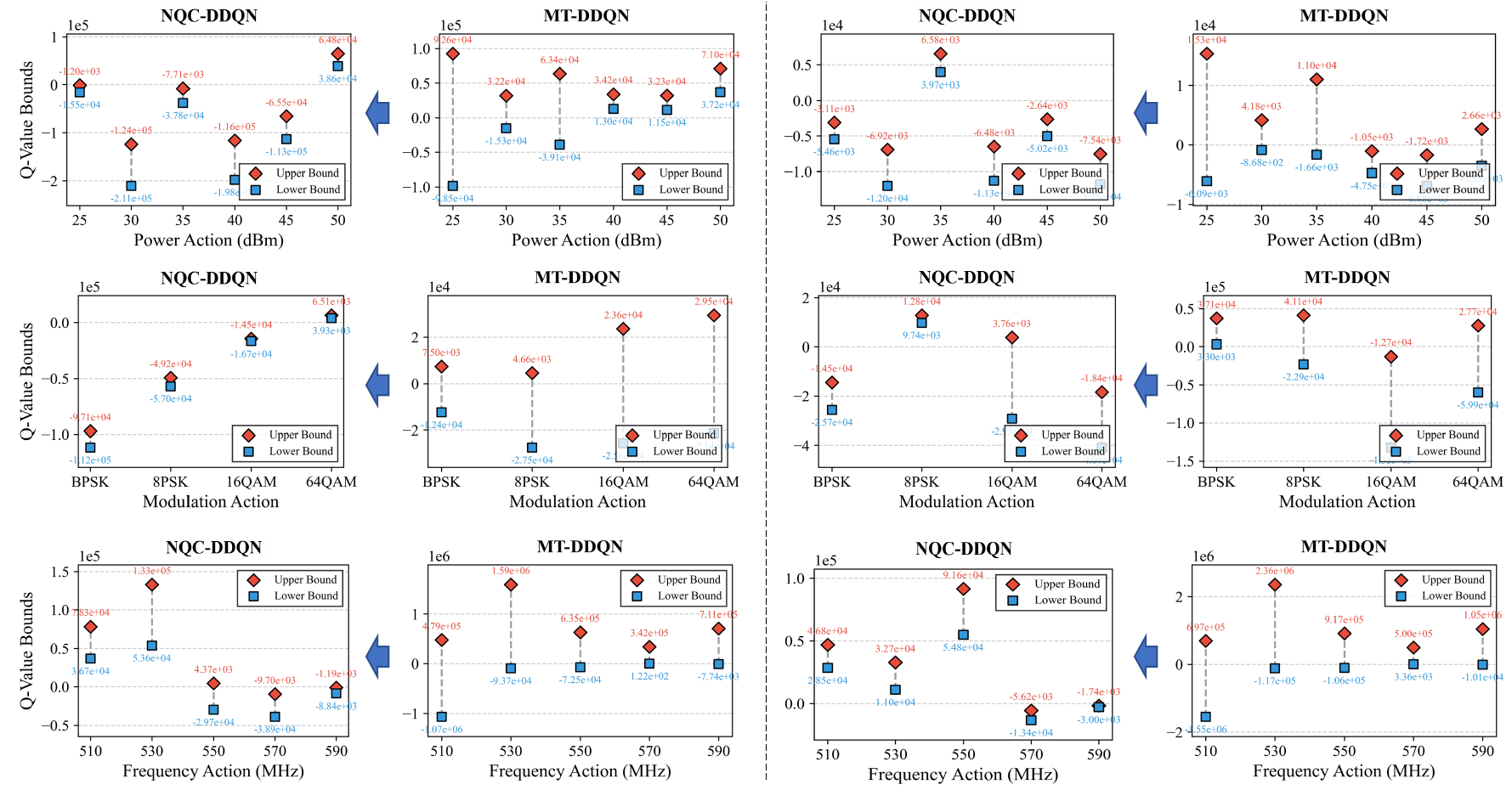}
	\caption{The output layer results of the power, modulation, and frequency networks. Two time slots are randomly chosen within their respective timescales.}
	\label{see_q}
\end{figure*}

To deeply analyze the decision stability of the NQC-DDQN algorithm and the reasons for partial decision deviations, we have visualized the decision-making process of the neural network at different timeslots under the condition of ${\varepsilon} = 10$ W, as illustrated in Fig. \ref{see_q}. The analysis reveals that the MT-DDQN algorithm, which lacks Q-value rectification, exhibits significant overlapping regions among Q-values of different actions, thereby inducing inconsistent action selection by the network. Notably, the NQC-DDQN algorithm not only adaptively reduces the range of the Q-value interval, but also ensures that the lower bound of the optimal action is higher than the upper bound of other actions. This ensures the consistency of the network output within the range of ${\varepsilon} < 10$. Finally, we analyze the causes of partial decision biases. During the execution phase, the agent has no prior knowledge of whether the input state is accurate. Therefore, it may erroneously output actions corresponding to true states if error-induced states exhibit approximate similarity to authentic state states. This is also a critical challenge that warrants further investigation in subsequent research.

\section{Conclusion}
This study addresses the decision-making failures of DRL methods in practical deployment caused by state uncertainty. Building upon the multi-timescale transmission model, we propose two effective solutions: the PGD-DDQN algorithm enhances output robustness by incorporating a regularization term to optimize neural network parameters under worst-case perturbations. The NQC-DDQN algorithm ensures policy stability through a nonlinear compression transformation mechanism for Q-value intervals, effectively eliminating action value aliasing in perturbation states. Statistical results indicate that the PGD-DDQN algorithm elevates the lower bound of performance, while the NQC-DDQN algorithm showcases remarkable robustness. These methodologies provide implementable solutions for DRL applications in uncertain real-world environments.

\section*{Appendix A}
\section*{Definition of $\psi ({v_z},{\beta _t})$}

\label{app:A}
\setcounter{equation}{0}
\renewcommand{\theequation}{A.\arabic{equation}}
In practical communication systems, accurate and reliable information recovery can be guaranteed when the bit error rate (BER) reaches $10^{-5}$. Under this BER constraint, distinct demodulation thresholds emerge for different modulation schemes. For two representative modulation formats – M-order Phase Shift Keying (M-PSK) and M-order Quadrature Amplitude Modulation (M-QAM) – their theoretical demodulation thresholds (SNR) can be mathematically derived through the following formulations
\begin{equation}
	\small
	\label{Q_function}
	Q(x) = \frac{1}{2}erfc(\frac{x}{{\sqrt 2 }})
\end{equation}
\begin{equation}
	\small
	\label{BER_PSK}
	\underbrace{\mathrm{BER} = \frac{2}{{{{\log }_2}M}}Q(\sqrt {2{{\log }_2}M \cdot \frac{{{E_b}}}{{{N_0}}}}  \cdot \sin (\frac{\pi }{M})),}_{\text{BER for M-PSK}}
\end{equation}
\begin{equation}
	\small
	\label{BER_QAM}
	\underbrace{\mathrm{BER} \approx \frac{4}{{{{\log }_2}M}}(1 - \frac{1}{{\sqrt M }})Q(\sqrt {\frac{{3{{\log }_2}M \cdot {E_b}/{N_0}}}{{M - 1}}} ),}_{\text{BER for M-QAM}}
\end{equation}
\begin{equation}
	\small
	\label{SJNR}
	SNR = {E_b}/{N_0} + 10{\log _{10}}({\log _2}M).
\end{equation}
Additionally, we have taken into account the coding gain provided by Low-density parity-check (LDPC) codes \cite{ref38}, which can further enhance the system performance. Based on the above discussion, we make appropriate assumptions and simplifications. When the ${\beta _t}$ at $t$ is less than the SNR of the selected modulation scheme $v_z$, $\psi ({v_z},{\beta _t}) = 0$. Under the condition that all other parameters are ideal, let $v_{\max }$ (with the highest modulation order among $\{ {v_1},{v_2}, \ldots ,{v_z}\}$) approach the Shannon limit, which means $\psi ({v_{\max }},{\beta _t}) = 1$. If ${\beta _t}$ is high and multiple modulation schemes are available, $\psi ({v_{\max }},{\beta _t})$ is calculated by $\Im = {\log _2}M \cdot (1 - BER)$. For example, if $v_{\max }$ is 64QAM, the BPSK scheme corresponds to $\psi ({v_{\max }},{\beta _t}) = \frac{{{{\log }_2}2 \cdot (1 - {{10}^{ - 5}})}}{{{{\log }_2}64 \cdot (1 - {{10}^{ - 5}})}} = \frac{1}{6}$.

\section*{Appendix B}
\section*{Proof of Lemma 1}

\label{app:B}
\setcounter{equation}{0}
\renewcommand{\theequation}{B.\arabic{equation}}
We prove the worst-case Bellman operator is a Contraction Mapping. Under the fixed policy $\pi _f$, for any two perturbed states ${\tilde s_1}$ and ${\tilde s_2}$, corresponding to $\tilde V_{{{\tilde s}_1}}^{{\pi _f}}(s)$ and $\tilde V_{{{\tilde s}_2}}^{{\pi _f}}(s)$, respectively, the following holds
\begin{equation}
	\small
	\label{s1}
	\mathcal{T} \tilde V_{{{\tilde s}_1}}^{{\pi _f}}(s) = \mathop {\min }\limits_{{{\tilde s}_1}} \sum\limits_{a \in {A_f}} {{\pi _f}(a|{{\tilde s}_1})} \sum\limits_{s' \in {S_f}} {p(s'|s,a)}  \cdot [r(s,a) + \gamma \tilde V_{{{\tilde s}_1}}^{{\pi _f}}(s')],
\end{equation}
\begin{equation}
	\small
	\label{s2}
	\mathcal{T} \tilde V_{{{\tilde s}_2}}^{{\pi _f}}(s) = \mathop {\min }\limits_{{{\tilde s}_2}} \sum\limits_{a \in {A_f}} {{\pi _f}(a|{{\tilde s}_2})} \sum\limits_{s' \in {S_f}} {p(s'|s,a)}  \cdot [r(s,a) + \gamma \tilde V_{{{\tilde s}_2}}^{{\pi _f}}(s').
 \end{equation}

${\tilde s_1}$ is the perturbed state that minimizes $\tilde V_{{{\tilde s}_1}}^{{\pi _f}}(s)$, we have \eqref{gais1}. By combining Equations \eqref{gais1} and \eqref{s2}, we derive \eqref{prove}.
\begin{equation}
	\small
	\label{gais1}
	\mathcal{T} \tilde V_{{{\tilde s}_1}}^{{\pi _f}}(s) \le \mathop {\min }\limits_{{{\tilde s}_2}} \sum\limits_{a \in {A_f}} {{\pi _f}(a|{{\tilde s}_2})} \sum\limits_{s' \in {S_f}} {p(s'|s,a)}  \cdot [r(s,a) + \gamma \tilde V_{{{\tilde s}_1}}^{{\pi _f}}(s')],
\end{equation}
\begin{equation}
	\small
	\label{prove}
	\begin{array}{l}
		\mathcal{T} \tilde V_{{{\tilde s}_1}}^{{\pi _f}}(s) - \mathcal{T} \tilde V_{{{\tilde s}_2}}^{{\pi _f}}(s)\\
		\le \mathop {\min }\limits_{{{\tilde s}_2}} \sum\limits_{a \in {A_f}} {{\pi _f}(a|{{\tilde s}_2})} \sum\limits_{s' \in {S_f}} {p(s'|s,a)}  \cdot [r(s,a) + \gamma \tilde V_{{{\tilde s}_1}}^{{\pi _f}}(s')]\\
		- \mathop {\min }\limits_{{{\tilde s}_2}} \sum\limits_{a \in {A_f}} {{\pi _f}(a|{{\tilde s}_2})} \sum\limits_{s' \in {S_f}} {p(s'|s,a)}  \cdot [r(s,a) + \gamma \tilde V_{{{\tilde s}_2}}^{{\pi _f}}(s')]\\
		\le \gamma  \cdot \mathop {\max }\limits_{{{\tilde s}_2}} \sum\limits_{a \in {A_f}} {{\pi _f}(a|{{\tilde s}_2})} \sum\limits_{s' \in {S_f}} {p(s'|s,a)}  \cdot [\tilde V_{{{\tilde s}_1}}^{{\pi _f}}(s') - \tilde V_{{{\tilde s}_2}}^{{\pi _f}}(s')]\\
		\le \gamma  \cdot \mathop {\max }\limits_{{{\tilde s}_2}} \sum\limits_{a \in {A_f}} {{\pi _f}(a|{{\tilde s}_2})} \sum\limits_{s' \in {S_f}} {p(s'|s,a)}  \cdot {\left\| {\tilde V_{{{\tilde s}_1}}^{{\pi _f}}(s') - \tilde V_{{{\tilde s}_2}}^{{\pi _f}}(s')} \right\|_\infty }\\
		= \gamma  \cdot {\left\| {\tilde V_{{{\tilde s}_1}}^{{\pi _f}}(s') - \tilde V_{{{\tilde s}_2}}^{{\pi _f}}(s')} \right\|_\infty }
	\end{array}.
\end{equation}

Then according to the Banach fixed-point theorem, $\tilde V_{\tilde s}^{{\pi _f}}(s)$ converges to the unique fixed point $\tilde V_{{{\tilde s}^*}}^{{\pi _f}}(s)$, which corresponds to the strongest perturbed state ${\tilde s^*}$.

\section*{Appendix C}
\section*{Proof of output boundedness}

\label{app:C}
\setcounter{equation}{0}
\renewcommand{\theequation}{C.\arabic{equation}}
We consider a $\chi $-layer fully connected neural network, where each layer is parameterized by a weight matrix ${\mathbf{W}^{(\chi )}}$ and a bias vector ${\mathbf{b}^{(\chi )}}$, with rectified linear unit (ReLU) activation functions. The input is bounded within an interval ${\mathbf{S}_{\varepsilon'} } = [({\underline{s}_1},{\bar s_1}),({\underline{s}_2},{\bar s_2}),...,({\underline{s}_n},{\bar s_n})]$, satisfying ${\underline{s}_n} \le {\bar s_n},\;\forall n$.

When $\chi  = 0$ corresponds to the input layer, ${\mathbf{x}^{(0)}} \in {\mathbf{S}_{\varepsilon'} }$, we have $\underline{\mathbf{x}}_n^{(\chi )} \le \bar {\mathbf{x}}_n^{(\chi )},\;\forall n$ by definition.
Assume the $\chi$-th layer contains ${d_\chi }$ neurons, with output intervals $[\underline{\mathbf{x}}_i^{(\chi )},\bar {\mathbf{x}}_i^{(\chi )}]$ satisfying $\underline{\mathbf{x}}_i^{(\chi )} \le \bar {\mathbf{x}}_i^{(\chi )},\;\forall i \in {d_\chi }$. Let the $({\chi}+1)$-th layer have ${d_{\chi  + 1}}$ neurons, with its output interval computed as follows

\textbullet\hspace{0.3em} \textit{Linear transformation (fully connected layer)}
\begin{equation}
	\small
	\label{fulllayer1}
	\underline{z}_j^{(\chi  + 1)} = \sum\limits_{i = 1}^{{d_\chi }} {W_{ji}^{(\chi )}} \cdot \underline{x}_i^{(\chi )} + b_j^{(\chi )},\;i \in {d_\chi },j \in {d_{\chi  + 1}},
\end{equation}
\begin{equation}
	\small
	\label{fulllayer2}
	\bar z_j^{(\chi  + 1)} = \sum\limits_{i = 1}^{{d_\chi }} {W_{ji}^{(\chi )}} \cdot \bar x_i^{(\chi )} + b_j^{(\chi )},\;i \in {d_\chi },j \in {d_{\chi  + 1}}.
\end{equation}

By term-wise comparison of the summation components, we obtain: $\underline{z}_j^{(\chi  + 1)} \le \bar z_j^{(\chi  + 1)},\;\forall j \in {d_{\chi  + 1}}$.

\textbullet\hspace{0.3em} \textit{Nonlinear transformation (ReLU)}
\begin{equation}
	\small
	\label{relu1}
	\underline{x}_j^{(\chi  + 1)} = \sigma (\underline{z}_j^{(\chi  + 1)}) = \max (0,\underline{z}_j^{(\chi  + 1)}),
\end{equation}
\begin{equation}
	\small
	\label{relu2}
	\bar x_j^{(\chi  + 1)} = \sigma (\bar z_j^{(\chi  + 1)}) = \max (0,\bar z_j^{(\chi  + 1)}).
\end{equation}

Since $\sigma ( \cdot )$ is monotonically increasing and $\underline{z}_j^{(\chi  + 1)} \le \bar z_j^{(\chi  + 1)}$, it follows that $\underline{x}_j^{(\chi  + 1)} \le \bar x_j^{(\chi  + 1)},\;\forall j \in {d_{\chi  + 1}}$. 

By mathematical induction, the output intervals of all layers remain bounded. Consequently, each dimension of the output layer satisfies ${\underline{Q}_y} \le {\bar Q_y},\;\forall y$.

\bibliographystyle{IEEEtran}
\bibliography{hq_tccn}

@ARTICLE{ref1,
  author={Pirayesh, Hossein and Zeng, Huacheng},
  journal={IEEE Communications Surveys \& Tutorials}, 
  title={Jamming Attacks and Anti-Jamming Strategies in Wireless Networks: A Comprehensive Survey}, 
  year={2022},
  volume={24},
  number={2},
  pages={767-809},
  doi={10.1109/COMST.2022.3159185}}

@ARTICLE{ref2,
  author={Khalek, Nada Abdel and Tashman, Deemah H. and Hamouda, Walaa},
  journal={IEEE Communications Surveys \& Tutorials}, 
  title={Advances in Machine Learning-Driven Cognitive Radio for Wireless Networks: A Survey}, 
  year={2024},
  volume={26},
  number={2},
  pages={1201-1237},
  doi={10.1109/COMST.2023.3345796}}

@ARTICLE{ref3_1,
  author={Hou, Zhifeng and Huang, Yuzhen and Chen, Jin and Li, Guoxin and Guan, Xinrong and Xu, Yifan and Chen, Runfeng and Xu, Yuhua},
  journal={IEEE Internet of Things Journal}, 
  title={Joint IRS Selection and Passive Beamforming in Multiple IRS-UAV-Enhanced Anti-Jamming D2D Communication Networks}, 
  year={2023},
  volume={10},
  number={22},
  pages={19558-19569},
  doi={10.1109/JIOT.2023.3281608}}

@ARTICLE{ref3_2,
  author={Li, Zan and Shi, Jia and Wang, Chao and Wang, Danyang and Li, Xiaomeng and Liao, Xiaomin},
  journal={China Communications}, 
  title={Intelligent covert communication design for cooperative cognitive radio network}, 
  year={2023},
  volume={20},
  number={7},
  pages={122-136},
  doi={10.23919/JCC.fa.2022-0514.202307}}

@ARTICLE{ref3,
  author={Han, Hao and Xu, Yuhua and Li, Wen and Wang, Ximing and Xu, Yifan and Zhang, Xiaokai and Gao, Yong},
  journal={IEEE Internet of Things Journal}, 
  title={Robust Spectrum Access Scheme Against Diverse Jamming Policies: A Prioritized Fictitious Rival-Play-Based Approach}, 
  year={2025},
  volume={12},
  number={1},
  pages={1-17},
  doi={10.1109/JIOT.2024.3459935}}

@ARTICLE{ref4,
  author={Amuru, SaiDhiraj and Dhillon, Harpreet S. and Buehrer, R. Michael},
  journal={IEEE Transactions on Wireless Communications}, 
  title={On Jamming Against Wireless Networks}, 
  year={2017},
  volume={16},
  number={1},
  pages={412-428},
  doi={10.1109/TWC.2016.2624291}}

@ARTICLE{ref5,
  author={Shi, Yuxin and Lu, Xinjin and An, Kang and Li, Yusheng and Zheng, Gan},
  journal={IEEE Internet of Things Journal}, 
  title={Efficient Index-Modulation-Based FHSS: A Unified Anti-Jamming Perspective}, 
  year={2024},
  volume={11},
  number={2},
  pages={3458-3472},
  doi={10.1109/JIOT.2023.3296605}}

@ARTICLE{ref6,
  author={Wang, Ximing and Wang, Jinlong and Xu, Yuhua and Chen, Jin and Jia, Luliang and Liu, Xin and Yang, Yijun},
  journal={IEEE Communications Magazine}, 
  title={Dynamic Spectrum Anti-Jamming Communications: Challenges and Opportunities}, 
  year={2020},
  volume={58},
  number={2},
  pages={79-85},
  doi={10.1109/MCOM.001.1900530}}

@ARTICLE{ref7,
  author={Zhang, Jianshu and Wu, Xiaofu and Tian, Feng},
  journal={IEEE Internet of Things Journal}, 
  title={Broadband Anti-Jamming With Distributed Sensing and Deep Reinforcement Learning: Spectrum Compression and Reward Estimation}, 
  year={2025},
  volume={12},
  number={2},
  pages={2203-2218},
  doi={10.1109/JIOT.2024.3469159}}

@ARTICLE{ref8,
  author={Si, Jiangbo and Huang, Rui and Li, Zan and Hu, Hang and Jin, Yuntao and Cheng, Julian and Al-Dhahir, Naofal},
  journal={IEEE Network}, 
  title={When Spectrum Sharing in Cognitive Networks Meets Deep Reinforcement Learning: Architecture, Fundamentals, and Challenges}, 
  year={2024},
  volume={38},
  number={1},
  pages={187-195},
  doi={10.1109/MNET.130.2200390}}

@ARTICLE{ref9,
  author={Pourranjbar, Ali and Kaddoum, Georges and Ferdowsi, Aidin and Saad, Walid},
  journal={IEEE Transactions on Communications}, 
  title={Reinforcement Learning for Deceiving Reactive Jammers in Wireless Networks}, 
  year={2021},
  volume={69},
  number={6},
  pages={3682-3697},
  doi={10.1109/TCOMM.2021.3062854}}

@ARTICLE{ref10,
  author={Janiar, Siavash Barqi and Wang, Ping},
  journal={IEEE Transactions on Vehicular Technology}, 
  title={Intelligent Anti-Jamming Based on Deep Reinforcement Learning and Transfer Learning}, 
  year={2024},
  volume={73},
  number={6},
  pages={8825-8834},
  doi={10.1109/TVT.2024.3359426}}

@ARTICLE{ref11,
  author={Li, Yangyang and Xu, Yuhua and Li, Guoxin and Gong, Yuping and Liu, Xin and Wang, Hao and Li, Wen},
  journal={IEEE Transactions on Information Forensics and Security}, 
  title={Dynamic Spectrum Anti-Jamming Access With Fast Convergence: A Labeled Deep Reinforcement Learning Approach}, 
  year={2023},
  volume={18},
  pages={5447-5458},
  doi={10.1109/TIFS.2023.3307950}}

@ARTICLE{ref12,
  author={Jia, Luliang and Qi, Nan and Su, Zhe and Chu, Feihuang and Fang, Shengliang and Wong, Kai-Kit and Chae, Chan-Byoung},
  journal={IEEE Communications Surveys \& Tutorials}, 
  title={Game Theory and Reinforcement Learning for Anti-jamming Defense in Wireless Communications: Current Research, Challenges, and Solutions}, 
  year={2024},
  doi={10.1109/COMST.2024.3482973}}

@ARTICLE{ref13,
  author={Qi, Jie and Zhang, Hongming and Qi, Xiaolei and Peng, Mugen},
  journal={IEEE Transactions on Vehicular Technology}, 
  title={Deep Reinforcement Learning Based Hopping Strategy for Wideband Anti-Jamming Wireless Communications}, 
  year={2024},
  volume={73},
  number={3},
  pages={3568-3579},
  doi={10.1109/TVT.2023.3324387}}

@ARTICLE{ref14,
  author={Cheng, Sixi and Ling, Xiang and Zhu, Lidong},
  journal={IEEE Open Journal of the Communications Society}, 
  title={Deep Reinforcement Learning-Based Anti-Jamming Approach for Fast Frequency Hopping Systems}, 
  year={2025},
  volume={6},
  number={},
  pages={961-971},
  doi={10.1109/OJCOMS.2025.3529982}}

@ARTICLE{ref15,
  author={Li, Wen and Qin, Yuan and Feng, Zhibin and Han, Hao and Chen, Jin and Xu, Yuhua},
  journal={IEEE Wireless Communications Letters}, 
  title={“Advancing Secretly by an Unknown Path”: A Reinforcement Learning-Based Hidden Strategy for Combating Intelligent Reactive Jammer}, 
  year={2022},
  volume={11},
  number={7},
  pages={1320-1324},
  doi={10.1109/LWC.2022.3165633}}

@ARTICLE{ref16,
  author={Liang, Fei and Shen, Cong and Yu, Wei and Wu, Feng},
  journal={IEEE Transactions on Communications}, 
  title={Towards Optimal Power Control via Ensembling Deep Neural Networks}, 
  year={2020},
  volume={68},
  number={3},
  pages={1760-1776},
  doi={10.1109/TCOMM.2019.2957482}}

@ARTICLE{ref17,
  author={Li, Xiangchen and Chen, Jienan and Ling, Xiang and Wu, Tingyong},
  journal={IEEE Transactions on Wireless Communications}, 
  title={Deep Reinforcement Learning-Based Anti-Jamming Algorithm Using Dual Action Network}, 
  year={2023},
  volume={22},
  number={7},
  pages={4625-4637},
  doi={10.1109/TWC.2022.3227575}}

@ARTICLE{ref18,
  author={Zhou, Quan and Niu, Yingtao and Xiang, Wanyu and Zhao, Liping},
  journal={IEEE Transactions on Industrial Informatics}, 
  title={A Novel Reinforcement Learning Algorithm Based on Broad Learning System for Fast Communication Antijamming}, 
  year={2025},
  volume={21},
  number={3},
  pages={2590-2599},
  doi={10.1109/TII.2024.3514085}}

@ARTICLE{ref20,
  author={Zhao, Haoqin and Si, Jiangbo and Li, Zan and Wang, Xiaoting and Al-Dhahir, Naofal},
  journal={IEEE Communications Letters}, 
  title={A Multi-Timescale Cross-Layer Anti-Jamming Scheme Under Rule Guidance}, 
  year={2025},
  volume={29},
  number={2},
  pages={259-263},
  doi={10.1109/LCOMM.2024.3510878}}

@ARTICLE{ref20_1,
  author={Wang, Shixiong and Dai, Wei and Sun, Jianyong and Xu, Zongben and Li, Geoffrey Ye},
  journal={IEEE Communications Magazine}, 
  title={Uncertainty Awareness in Wireless Communications and Sensing}, 
  year={2025},
  volume={},
  number={},
  pages={1-9},
  doi={10.1109/MCOM.001.2400714}}

@ARTICLE{ref20_1_1,
  author={Wu, Yuhang and Zhou, Fuhui and Wu, Wei and Wu, Qihui and Ng, Derrick Wing Kwan and Quek, Tony Q. S.},
  journal={IEEE Transactions on Wireless Communications}, 
  title={Robust Resource Allocation for RSMA Spectrum Sharing Networks}, 
  year={2024},
  volume={23},
  number={11},
  pages={16375-16389},
  doi={10.1109/TWC.2024.3440652}}

@ARTICLE{ref20_1_2,
  author={Zhou, Gui and Pan, Cunhua and Ren, Hong and Wang, Kezhi and Nallanathan, Arumugam},
  journal={IEEE Transactions on Signal Processing}, 
  title={A Framework of Robust Transmission Design for IRS-Aided MISO Communications With Imperfect Cascaded Channels}, 
  year={2020},
  volume={68},
  number={},
  pages={5092-5106},
  doi={10.1109/TSP.2020.3019666}}

@INPROCEEDINGS{ref20_1_3,
  author={Zhang, Hanwen and Sun, Haijian and He, Tianyi and Xiang, Weiming and Hu, Rose Qingyang},
  booktitle={2024 IEEE International Conference on Communications Workshops (ICC Workshops)}, 
  title={Energy Efficient Robust Beamforming for Vehicular ISAC with Imperfect Channel Estimation}, 
  year={2024},
  volume={},
  number={},
  pages={1864-1869},
  doi={10.1109/ICCWorkshops59551.2024.10615653}}

@article{ref21,
title = {Improving robustness by action correction via multi-step maximum risk estimation},
journal = {Neural Networks},
volume = {184},
pages = {107045},
year = {2025},
issn = {0893-6080},
doi = {https://doi.org/10.1016/j.neunet.2024.107045},
url = {https://www.sciencedirect.com/science/article/pii/S0893608024009742},
author = {Qinglong Chen and Kun Ding and Xiaoxiong Zhang and Hui Zhang and Fei Zhu},
}

@inproceedings{ref22,
  title={A review of uncertainty for deep reinforcement learning},
  author={Lockwood, Owen and Si, Mei},
  booktitle={Proceedings of the AAAI Conference on Artificial Intelligence and Interactive Digital Entertainment},
  volume={18},
  number={1},
  pages={155-162},
  year={2022}
}

@article{ref23,
  title={Risk-averse model uncertainty for distributionally robust safe reinforcement learning},
  author={Queeney, James and Benosman, Mouhacine},
  journal={Advances in Neural Information Processing Systems},
  volume={36},
  pages={1659--1680},
  year={2023}
}

@article{ref24,
  title={Robust deep reinforcement learning against adversarial perturbations on state observations},
  author={Zhang, Huan and Chen, Hongge and Xiao, Chaowei and Li, Bo and Liu, Mingyan and Boning, Duane and Hsieh, Cho-Jui},
  journal={Advances in Neural Information Processing Systems},
  volume={33},
  pages={21024--21037},
  year={2020}
}

@article{ref25,
  title={Robust deep reinforcement learning through adversarial loss},
  author={Oikarinen, Tuomas and Zhang, Wang and Megretski, Alexandre and Daniel, Luca and Weng, Tsui-Wei},
  journal={Advances in Neural Information Processing Systems},
  volume={34},
  pages={26156--26167},
  year={2021}
}

@ARTICLE{ref26,
  author={Li, Wen and Xu, Yuhua and Chen, Jin and Yuan, Hongcheng and Han, Hao and Xu, Yifan and Feng, Zhibin},
  journal={IEEE Wireless Communications Letters}, 
  title={Know Thy Enemy: An Opponent Modeling-Based Anti-Intelligent Jamming Strategy Beyond Equilibrium Solutions}, 
  year={2023},
  volume={12},
  number={2},
  pages={217-221},
  doi={10.1109/LWC.2022.3219434}}

@ARTICLE{ref27,
  author={Bai, Qinbo and Wang, Jintao and Zhang, Yue and Song, Jian},
  journal={IEEE Transactions on Cognitive Communications and Networking}, 
  title={Deep Learning-Based Channel Estimation Algorithm Over Time Selective Fading Channels}, 
  year={2020},
  volume={6},
  number={1},
  pages={125-134},
  doi={10.1109/TCCN.2019.2943455}}

@ARTICLE{ref28,
  author={Li, Wen and Chen, Jin and Liu, Xin and Wang, Ximing and Li, Yangyang and Liu, Dianxiong and Xu, Yuhua},
  journal={IEEE Wireless Communications}, 
  title={Intelligent Dynamic Spectrum Anti-Jamming Communications: A Deep Reinforcement Learning Perspective}, 
  year={2022},
  volume={29},
  number={5},
  pages={60-67},
  doi={10.1109/MWC.103.2100365}}

@ARTICLE{ref29,
  author={Xu, Jianliang and Lou, Huaxun and Zhang, Weifeng and Sang, Gaoli},
  journal={IEEE Access}, 
  title={An Intelligent Anti-Jamming Scheme for Cognitive Radio Based on Deep Reinforcement Learning}, 
  year={2020},
  volume={8},
  number={},
  pages={202563-202572},
  doi={10.1109/ACCESS.2020.3036027}}

@ARTICLE{ref30,
  author={Zhang, Yupei and Zhao, Zhijin and Zheng, Shilian and Qiang, Fangfang},
  journal={IEEE Transactions on Cognitive Communications and Networking}, 
  title={Intelligent Anti-Jamming Decision With Continuous Action and State in Bivariate Frequency Agility Communication System}, 
  year={2023},
  volume={9},
  number={6},
  pages={1579-1595},
  doi={10.1109/TCCN.2023.3306363}}

@ARTICLE{ref31,
  author={Bai, Zixuan and Shi, Jia and Li, Zan and Li, Meng and Chen, Kwang-Cheng},
  journal={IEEE Transactions on Cognitive Communications and Networking}, 
  title={Rule-Guided DRL for UAV-Assisted Wireless Sensor Networks With No-Fly Zones Safety}, 
  year={2025},
  volume={11},
  number={2},
  pages={1268-1280},
  doi={10.1109/TCCN.2024.3460759}}

@inproceedings{ref32,
  title={Deep reinforcement learning with double q-learning},
  author={Van Hasselt, Hado and Guez, Arthur and Silver, David},
  booktitle={Proceedings of the AAAI conference on artificial intelligence},
  volume={30},
  number={1},
  year={2016}
}

@article{ref33,
  title={Interpreting Adversarial Attacks and Defences using Architectures with Enhanced Interpretability},
  author={Rao, Akshay G and Lakshminarayanan, Chandrashekhar and Rajkumar, Arun},
  journal={arXiv preprint arXiv:2502.15017},
  year={2025}
}

@inproceedings{ref34,
  title={Scalable verified training for provably robust image classification},
  author={Gowal, Sven and Dvijotham, Krishnamurthy Dj and Stanforth, Robert and Bunel, Rudy and Qin, Chongli and Uesato, Jonathan and Arandjelovic, Relja and Mann, Timothy and Kohli, Pushmeet},
  booktitle={Proceedings of the IEEE/CVF International Conference on Computer Vision},
  pages={4842--4851},
  year={2019}
}

@inproceedings{ref35,
  title={Safe reinforcement learning via shielding},
  author={Alshiekh, Mohammed and Bloem, Roderick and Ehlers, R{\"u}diger and K{\"o}nighofer, Bettina and Niekum, Scott and Topcu, Ufuk},
  booktitle={Proceedings of the AAAI conference on artificial intelligence},
  volume={32},
  number={1},
  year={2018}
}

@article{ref36,
  title={Anti-jamming communications using spectrum waterfall: A deep reinforcement learning approach},
  author={Liu, Xin and Xu, Yuhua and Jia, Luliang and Wu, Qihui and Anpalagan, Alagan},
  journal={IEEE Communications Letters},
  volume={22},
  number={5},
  pages={998--1001},
  year={2018},
  publisher={IEEE}
}

@ARTICLE{ref37,
  author={Xu, Kaidi and Van Huynh, Nguyen and Li, Geoffrey Ye},
  journal={IEEE Transactions on Communications}, 
  title={Distributed-Training-and-Execution Multi-Agent Reinforcement Learning for Power Control in HetNet}, 
  year={2023},
  volume={71},
  number={10},
  pages={5893-5903},
  doi={10.1109/TCOMM.2023.3300331}}

@ARTICLE{ref38,
  author={Zhu, Min and Guo, Quan and Bai, Baoming and Ma, Xiao},
  journal={IEEE Transactions on Communications}, 
  title={Reliability-Based Joint Detection-Decoding Algorithm for Nonbinary LDPC-Coded Modulation Systems}, 
  year={2016},
  volume={64},
  number={1},
  pages={2-14},
  doi={10.1109/TCOMM.2015.2487454}}
\end{document}